\pdfoutput=1
\documentclass[11pt]{article}
\usepackage[preprint]{acl}

\usepackage{times}
\usepackage{latexsym}
\usepackage{amsmath,amssymb,amsfonts}
\usepackage{graphicx}
\usepackage{booktabs}
\usepackage{longtable}
\usepackage{caption}
\usepackage{stfloats}
\usepackage{multirow} 
\usepackage{comment}
\usepackage{mfirstuc,textcase,xspace,graphicx}
\usepackage{longtable}
\usepackage{wrapfig}
\usepackage{comment}  
\usepackage{algorithm}
\usepackage{algorithmic}
\usepackage{float}
\usepackage{xcolor,colortbl} 
\colorlet{red}{black}
\usepackage{amsmath,amsfonts} %
\usepackage{graphicx}%
\usepackage{textcomp}%
\usepackage{array}%
\usepackage{pifont}%
\usepackage{amsthm}%
\usepackage{bm}
\usepackage{setspace}%
\usepackage{multirow}%
\usepackage{siunitx}
\usepackage{footnote}%
\usepackage[protrusion=true,expansion=true,activate={true,nocompatibility},final,tracking=true,kerning=true,spacing=true,stretch=10,shrink=10,disable=footnote]{microtype}
\usepackage{blindtext}
\usepackage{ctable} 
\usepackage{pifont}
\usepackage{booktabs}
\usepackage{makecell}
\usepackage{tabularx}
\usepackage{enumitem}
\graphicspath{ {image/} }
\usepackage{titlesec}
\usepackage{multirow}
\newcommand{\ul}[1]{\underline{#1}}

\usepackage{microtype}
\usepackage{inconsolata}
\usepackage{url}
\usepackage{pifont}

\DeclareUnicodeCharacter{2009}{,}
\DeclareUnicodeCharacter{2248}{\ensuremath{\approx}}

\title{Lyapunov-Inspired LyRIC Activation and GLARE Attention in Chaos-Guided State Space Modeling for EMG-To-Speech (ETS) Synthesis}

\author{Sajid Fardin Dipto \And
Tarikul Islam Tamiti \And
Luke Baja-Ricketts \AND
David Vergano \And
Anomadarshi Barua \AND
\normalfont Department of Cyber Security Engineering \\
\normalfont George Mason University \\
\normalfont Fairfax, VA, USA}

\date{}

\begin{document}

\maketitle

\begin{abstract}
Electromyography-to-Speech (ETS) synthesis is typically a \textit{non-linear, chaotic} dynamical system. {\color{red}However, no prior work has studied the chaotic behavior of ETS synthesis to date.} Yet, prior works strictly rely on standard reconstruction metrics with parameter-heavy transformers that systematically over-smooth natural acoustic dynamics. {\color{red}To close this gap, for the first time, we propose a chaos-inspired Lyapunov-derived new activation function (LyRIC) with two novel chaotic loss functions -- Lyapunov Exponent Regularization and Multi-Scale Detrended Fluctuation Analysis -- to explicitly capture the deterministic chaos of human phonation.} In addition, we introduce a compressed novel encoder -- GLAME, which synergizes global Mamba state-space modeling with localized GLARE attention. {\color{red}We comprehensively do frame-level acoustic evaluation in a multi-language and multi-speaker setup on two datasets - English and Mandarin.} It outperforms the established baseline with a 4.69$\times$ increase in objective intelligibility (STOI: 0.61 vs. 0.13) and a 2.08$\times$ improvement in spectral reconstruction (LSD: 1.08 vs. 2.25). {\color{red}Importantly, this improvement is achieved with 73.49\% fewer parameters (14.34M vs. 54.10M), establishing a new baseline in the ETS synthesis task. To the best of our knowledge, this is the first work proving that integrating non-linear chaotic physics into neural networks yields superior but compact inductive biases for real-time ETS synthesis task.}
\end{abstract}

\section{Introduction and Related Work}
\label{sec:intro}

\begin{figure*}[t]
    \centering
    \includegraphics[width=1.0\textwidth]{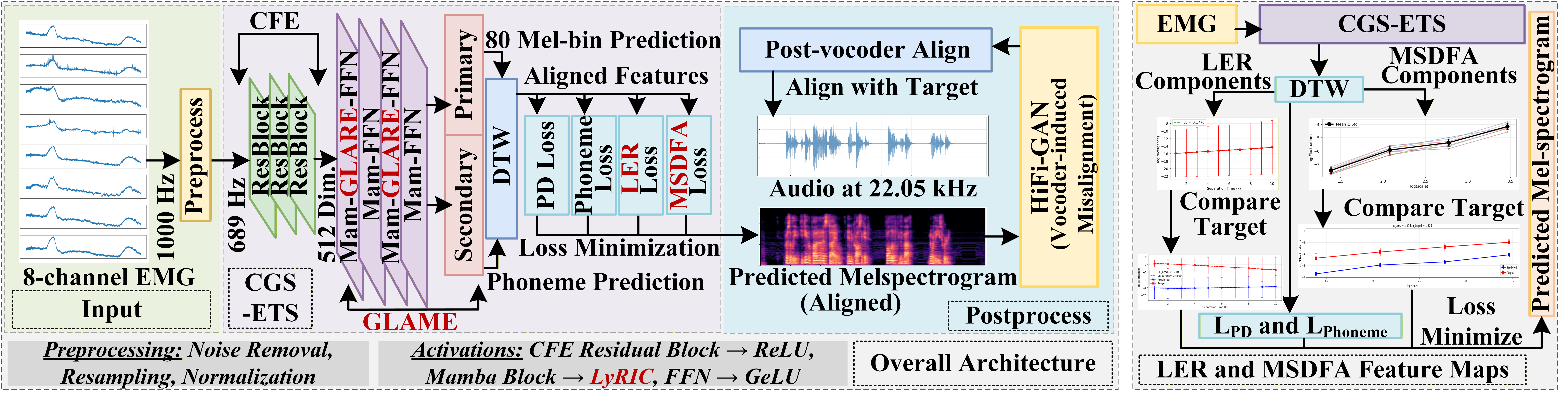}
    \caption{Detailed architecture of CGS-ETS with chaos-driven LyRIC activation and GLARE attention.}
    \label{fig:cgs-ets}
     \vspace{-1.0em}
\end{figure*}

{\color{black}

{\color{red}Silent speech interfaces (SSIs) synthesize unvocalized speech by decoding articulatory biosignals, assisting speech-impaired users. Electromyography (EMG) captures the muscle activity needed for this synthesis~\cite{dong2023, satterlee2025}.}

\textbf{What current ETS systems miss:} Despite steady improvements in Word Error Rate (WER), {\color{red}recent ETS models published in ACL and EMNLP} \cite{gaddy-klein-2020-digital, gaddy-klein-2021-improved} and \cite{scheck2023, ren2024, xie2025neural} suffer from the following three fundamental limitations.

{\color{red}\textbf{\textit{\ul{First}}}, approaching speech synthesis as a smoothed regression task optimizes for frame-level losses while neglecting the \textit{non-linear chaotic dynamics} of human phonation. Penalizing natural chaotic variations inevitably forces models to produce muffled, robotic prosody. \textbf{\textit{\ul{Second}}}, recent models \cite{gaddy-klein-2020-digital, gaddy-klein-2021-improved} rely on parameter-heavy Transformers (exceeding 44M parameters). This architectural bloat fails to exploit the localized, temporal shift-invariance of EMG signals, violating the low-latency constraints required for SSIs. \textbf{\textit{\ul{Third}}}, prior evaluations rely overwhelmingly on WER or CER, leaving the true perceptual fidelity, spectral distortion, and objective intelligibility of previous systems uncharacterized.}


\textit{To address these gaps, we propose \textbf{CGS-ETS (\underline{C}haos-\underline{G}uided \underline{S}tate Space Modeling for \underline{E}MG-\underline{T}o-\underline{S}peech)}, which is the first to unify non-linear chaos theory with efficient State Space Model (SSM), achieving better acoustic fidelity with a drastically reduced parameter footprint.} Our core contributions are manifold: {\color{red}
\textbf{\ul{(1)}} We invent and integrate \textbf{\textit{\underline{G}rouped \underline{L}ocal \underline{A}ttention with \underline{R}elative Positional \underline{E}mbeddings (GLARE)}} to the ETS domain. GLARE captures temporal shift-invariance while accelerating inference for long-context EMG sequences, providing an efficient alternative to standard Multi-Head Attention (MHA) for ETS synthesis.} \textbf{\ul{(2)}} We propose the \textbf{\textit{\underline{GLA}RE-fused \underline{M}amba \underline{E}ncoder (GLAME)}}, a 
parameter-efficient SSM architecture, where standard SiLU activations are replaced with {\color{red}our newly designed \textbf{\textit{\underline{Ly}apunov \underline{R}egularized 
\underline{I}mplicit \underline{C}urve}} (LyRIC) activation function to inject chaotic awareness into the state-space trajectory, achieving a 77.63\% reduced complexity.}
\textbf{\ul{(3)}} We introduce two novel optimization objectives to constrain the network to the deterministic chaotic invariants of speech: \textit{\textbf{LER}} \textbf{(Lyapunov Exponent Regularization)} loss penalizes mismatches in local divergence rates by aligning Lyapunov Exponents (LE) across varying temporal resolutions; and \textit{\textbf{MSDFA}} \textbf{(Multi-Scale Detrended Fluctuation Analysis)} quantifies the fractal-like, long-range temporal chaotic correlations strictly overlooked by standard spectrogram losses.
\textbf{\ul{(4)}} {\color{red} Under {\textbf{LER}} and {\textbf{MSDFA}} supervision, GLAME achieves $4.69\times$ higher intelligibility, $2.08\times$ better spectral reconstruction, and 73.49\% fewer parameters than recent baselines in multilingual settings.}} 

\section{Overall Architecture}
\label{sec:overallarchitecture}

We show CGS-ETS in Fig. \ref{fig:cgs-ets}. It takes 8-channel EMG data as input, preprocesses it, and generates melspectrogram (mel) bins, which are converted to audio at the last stage by a HiFi-GAN vocoder \cite{hifigan2020}. {\color{red}Our previously mentioned \textit{novel four innovations} in CGS-ETS -- an efficient alternative \textit{\textbf{GLARE}} to standard MHA, a first chaos-derived activation function \textbf{\textit{LyRIC}}, two chaos-guided loss components \textit{\textbf{LER}} and \textit{\textbf{MSDFA}}, and a \textit{very lightweight} encoder \textbf{\textit{GLAME}} are explained below sequentially in the rest of the paper.}

\subsection{Design of the GLARE}
\label{sec:GLARE}

\begin{figure}[t]
    \centering
    \includegraphics[width=0.48\textwidth]{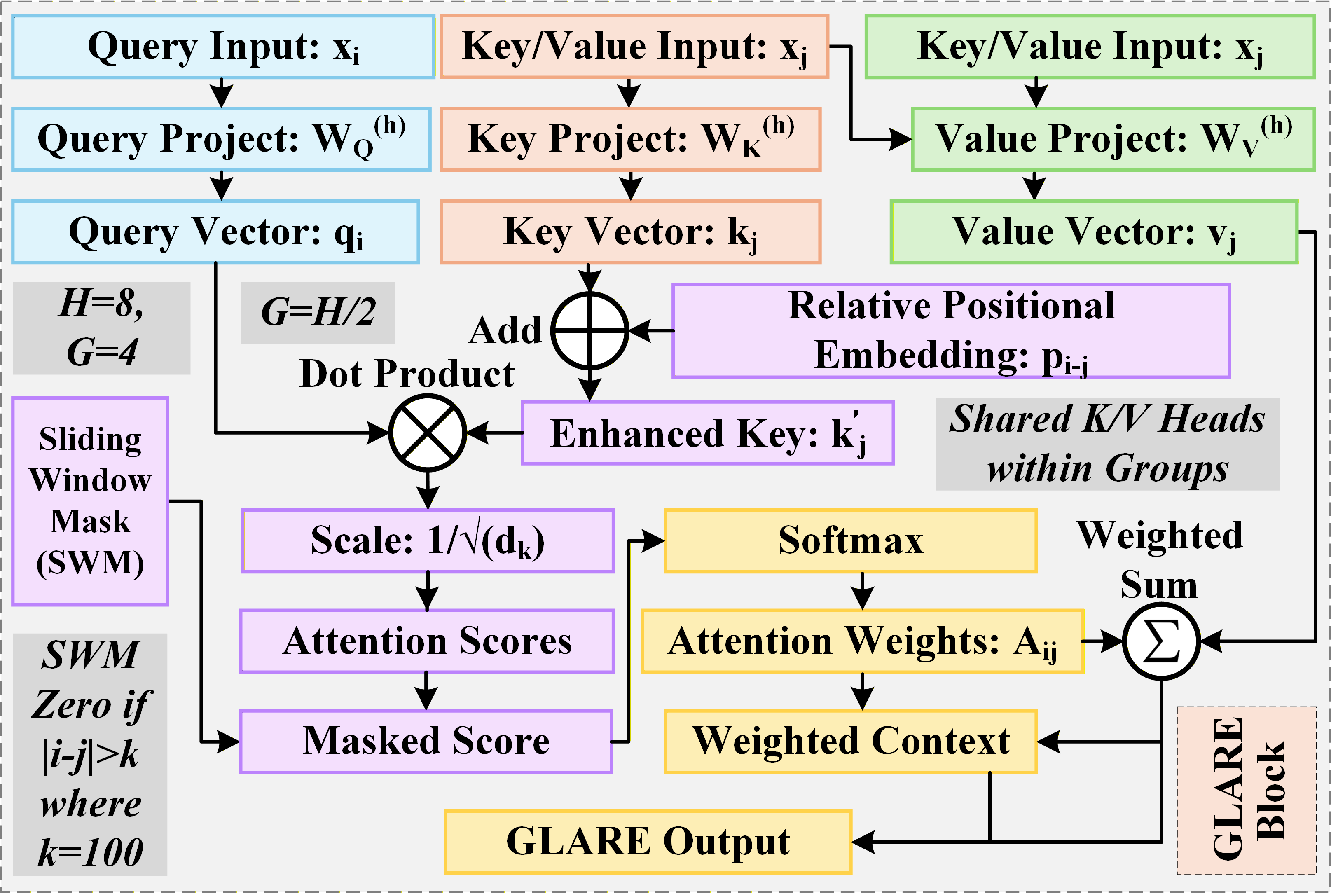}
    \caption{Architecture of GLARE attention.}
    \label{fig:glare}
  \vspace{-1.0em}
\end{figure}

Please note that the Grouped Query Attention (GQA) \cite{GQA2023} is \textit{a generalization of Multi-Query Attention (MQA)} \cite{MQA2019}, and Sliding Window Attention (SWA) captures fine-grained \textit{local context} by sliding over the input sequence \cite{swa2020} through a sliding window mask (SWM). We adopt GQA within the SWA module rather than employing standard MHA to reduce memory bandwidth overhead during inference \textit{(see Table \ref{tab:Swa_ablation} for ablation on different attention mechanisms)}. Furthermore, we deliberately integrate Relative Positional Embeddings (RPE) from \cite{gaddy-klein-2021-improved} to ensure the attention remains highly robust to natural temporal variations and the precise local alignments required for high-fidelity phonetic decoding. Thus, we propose GLARE, a unified attention mechanism that integrates SWM and RPE with GQA, as illustrated in Fig. \ref{fig:glare}. {\color{red}Like optical glare, which achieves maximum perceptual signal through a directed, concentrated beam rather than diffuse ambient light, our GLARE concentrates attention within a sliding local window, groups query computations for efficiency, and encodes positional relationships through relative embeddings rather than attending globally across all token positions, yielding competitive performance at substantially reduced computational cost compared to full MHA.}

\textbf{Governing Equation Design:} GLARE is shown in Fig. \ref{fig:glare}. Incorporating RPE, we compute the attention weight $A^{(h)}_{ij}$ for head $h \in \{0, \ldots, H{-}1\}$ between query position $i$ and key position $j$ as:

{\small
\begin{equation}
\vspace{-0.01em}
    A^{(h)}_{ij} = \frac{
        \left(W_K^{(g(h))} \mathbf{x}_j + \mathbf{p}_{i-j}\right)^\top 
        \left(W_Q^{(h)} \mathbf{x}_i\right)
    }{\sqrt{d_k}}
\end{equation} }

\noindent where {\small $g(h) = \left\lfloor \frac{h \cdot G}{H} \right\rfloor$} and $\mathbf{x}_i \in \mathbb{R}^{d_{model}}$ is the input feature vector at position $i$, {\small $W_Q^{(h)} \in \mathbb{R}^{d_k \times d_{model}}$} is the query projection matrix 
for head $h$, {\small $W_K^{(g(h))}$} and {\small $W_V^{(g(h))}$} are the shared key and value projection matrices for the key-value group assigned to head $h$, $\mathbf{p}_{i-j} \in \mathbb{R}^{d_k}$ is a learned relative position embedding indexed by the signed distance $i - j$, and $g(h)$ maps each of the $H$ query heads to one of $G$ shared key-value groups. For {\small $|i - j| > k$}, {\small $A^{(h)}_{ij}$} is set to zero, effectively restricting attention to a local window of $k$ steps in each direction. We keep $k=100$ and $H=8$. 
We partition the $H$ query heads into $G$ groups, where each group shares a single key and value head. GQA subsumes both MHA ($G=H$) and MQA ($G=1$) \cite{GQA2023}, and we set $G = \lfloor H/2 \rfloor$ in our implementation, halving the number of key-value heads relative to query heads, which reduces the KV cache size and alleviates the memory bandwidth bottleneck, particularly critical for the long-context EMG sequences in our setting.

\subsection{Design of the LyRIC Activation}
\label{subsec:lyric}

We mathematically derive the LyRIC activation curve through the expansion of the logarithmic divergence term proposed in the Lyapunov theory \cite{oseledec1968multiplicative, wolf1985determining}.  

\textbf{Logarithmic Divergence Implementation (\ul{see Alg. }\ref{alg:LER}):} The largest Lyapunov Exponent (LE) is computed on non-overlapping windows with time-delay embedding dimension $m$, and time delay $\tau$. The inputs are first embedded into a higher-dimensional phase space using time-delay embedding, $E_V[i] = [x[i, :], \dots, x[i + (m - 1)\tau]]$ where $x$ $\in$ $\mathbb{R}^{T \times C}$ is the input feature sequence with $T$ time frames and $C$ feature channels, and $d$ = $m \cdot C$ is the embedding dimension (\textbf{Rows \textcircled{\scriptsize 1}--\textcircled{\scriptsize 4}}). Then, the Pairwise Distance (PD) matrix is computed as $D[i, j]$ 
for $i, j \in \{0, 1, \ldots, M-1\}$. Here, the maximum number of embedding vectors to use is $M$ = $T-(m-1)\tau$ (\textbf{Rows \textcircled{\scriptsize 5}--\textcircled{\scriptsize 7}}). Then, the nearest neighbor for each point is calculated, $j^*(i) = \arg \min_{j \in \{1, \dots, M\}} D[i, j]$ while excluding temporally correlated neighbors using Theiler window \cite{Theiler1986} and track the distance evolution $d_k[i]$ for each time offset $k$ 
(\textbf{Rows \textcircled{\scriptsize 9}--\textcircled{\scriptsize 12}}). Then, \textbf{\textit{the logarithmic divergence}} of nearest neighbor pairs over time for each $k$ is estimated as, $S(k) = \frac{1}{N_k} \sum_{i=1}^{M} \log (d_k[i] + \epsilon_1)$, where $\epsilon_1$ prevents logarithm of zero (\textbf{Row \textcircled{\scriptsize 14}}). 

\textbf{Activation Function Design:} We approximate $S(k)$ with a singular logarithmic divergence term to seamlessly integrate this dynamical systems metric as an element-wise neural activation. We generalize the local state-space divergence, $d_k[i]$, to the absolute magnitude of the input activation $\vert x \vert$, which ensures a valid logarithmic domain ($\vert x \vert \geq 0$). Then, we introduce a bounded stability margin of $\epsilon_1 = 1$ to strictly prevent domain violations within the logarithm and avoid singularity collapse. Thus, $\epsilon_1$ offset guarantees zero-centered topological stability (i.e., $f(0) = 0$), making it suitable for stable gradient propagation in deep architectures. This mathematically motivated derivation yields the continuous, non-linear mapping $h(x) = \log(\vert x \vert + 1)$. We introduce a learnable scalar parameter $\alpha$ to instantiate this divergence metric as a highly adaptive neural activation that exponentiates the absolute magnitude to dynamically control the local state-space expansion rate. Then, we modulate the logarithmic term with the raw input $x$ to ensure odd symmetry (directional phase preservation), strictly bound the activation magnitude, and apply a denominator-based stability normalization. This theoretically motivated synthesis yields an intermediate function, defined as: 

\begin{equation}
    \small g(x) = x \left[ \frac{\log(\vert x \vert^\alpha + 1)}{\vert x \vert^\alpha + 1} \right]
    \label{eqn2}
\end{equation}

\begin{figure}[t]
    \centering
    \includegraphics[width=1\columnwidth, height=5cm, keepaspectratio]{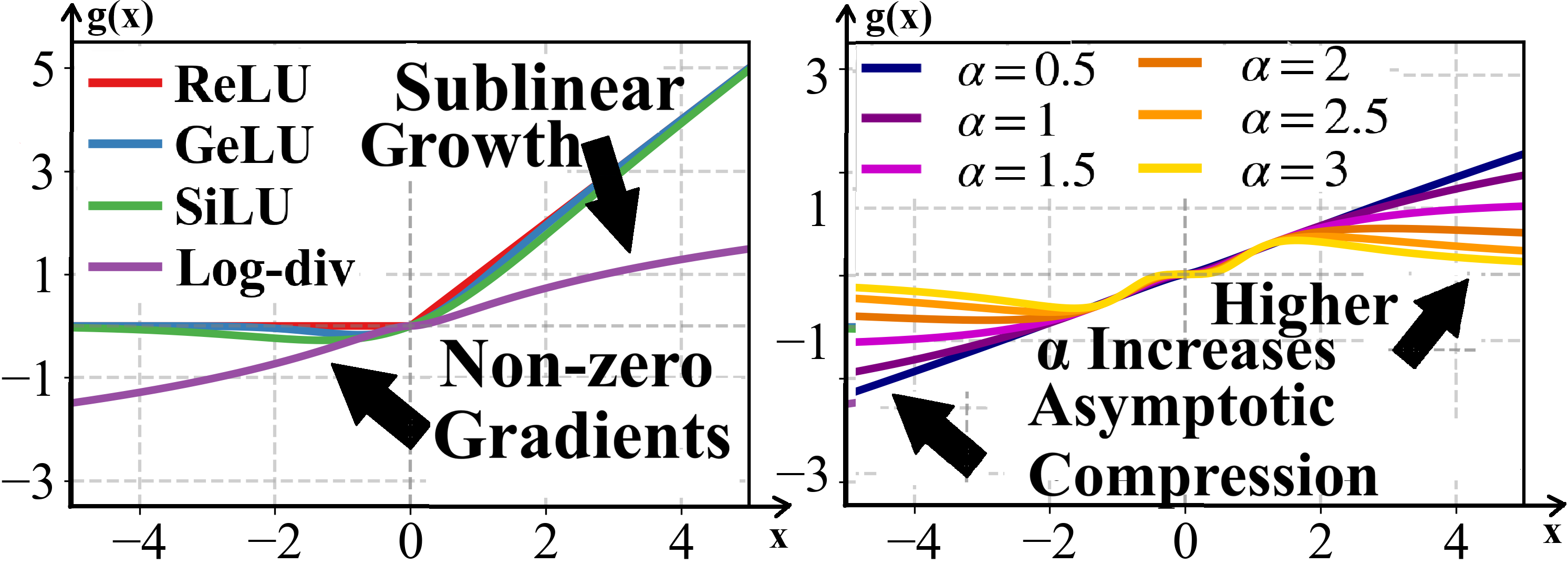}
    \caption{{\color{red}(Left) Comparing different activation functions (Eqn. ~\ref{eqn2}). (Right) Effect of $\alpha$ on log-divergence.}}
    \label{fig:log-div-plot}
    \vspace{-1.0em}
\end{figure}

This formulation is mathematically guaranteed to preserve critical topological and optimization properties \textit{(see 
Appendix~\ref{subsubsec:lyric_properties})}, as illustrated in Fig.~\ref{fig:log-div-plot}. However, the direct computation of Eqn. \ref{eqn2} introduces non-trivial inference latency due to an increase in computational overhead. In particular, the nested logarithmic and fractional evaluations demand considerably higher Floating Point Operations (FLOPs) compared to standard, hardware-optimized activations such as ReLU, GeLU, or SiLU \textit{(see computational profiling in Appendix \ref{subsubsec:log-divergence}, Table \ref{tab:flops_activation})}. Therefore, we construct a polynomial approximation through \textit{Maclaurin series and geometric series} expansions to mitigate this computational bottleneck while preserving the chaotic properties. The constituent logarithmic and rational terms from Eqn. \ref{eqn2} decompose as follows:

\vspace{-1.5em}
\begin{equation}
    \small \log(\vert x \vert^\alpha + 1) = \vert x \vert^\alpha - \frac{\vert x \vert^{2\alpha}}{2} + \frac{\vert x \vert^{3\alpha}}{3} - \dots
    \label{eqn3}
    \vspace{-2.2em}
\end{equation}

\begin{equation}
    \small \frac{1}{\vert x \vert^\alpha + 1} = 1 - \vert x \vert^\alpha + \vert x \vert^{2\alpha} - \vert x \vert^{3\alpha} + \dots
    \label{eqn4}
    \vspace{-0.0em}
\end{equation}


\noindent We derive a polynomial representation by computing the Cauchy product of Eqns. \ref{eqn3} and \ref{eqn4}, and subsequently multiplying the series by $x$. Then the term $g(x)$ in Eqn. \ref{eqn2} becomes:

\vspace{-00.8em}
\begin{equation}
    \small g(x)=x\vert x \vert^\alpha - \frac{3}{2}x\vert x \vert^{2\alpha} + \frac{11}{6}x\vert x \vert^{3\alpha} - \dots
    \label{eqn5}
    \vspace{-0.0em}
\end{equation}

\noindent We truncate the infinite series and isolate the leading-order expansion term, $x\vert x \vert^\alpha$, as our intention is to minimize computational overhead while preserving the essential non-linear dynamics of Eqn. \ref{eqn2}. {\color{black}This  approximation retains the foundational topological properties of the original Lyapunov derived function (Eqn. \ref{eqn2}), which are odd symmetry, origin linearity, and adaptive phase-space expansion \textit{(compare Fig. \ref{fig:log-div-plot} and \ref{fig:LyRIC_main})} without the prohibitive FLOPs burden of nested logarithmic evaluations. However, arbitrarily parameterizing the expansion coefficient $\alpha$ in $x\vert x \vert^\alpha$ can induce discontinuities \textit{(i.e., $\alpha \in \{-1,-1.5\}$; see Fig. \ref{fig:LyRIC_discont})} across the activation manifold. We therefore constrain $\alpha > 0$, which mathematically guarantees $C^1$ continuity (continuity at the shared junction between the two curves) at the operational points.

\begin{figure}[t]
    \centering
    \includegraphics[width=1\columnwidth, height=5cm, keepaspectratio]{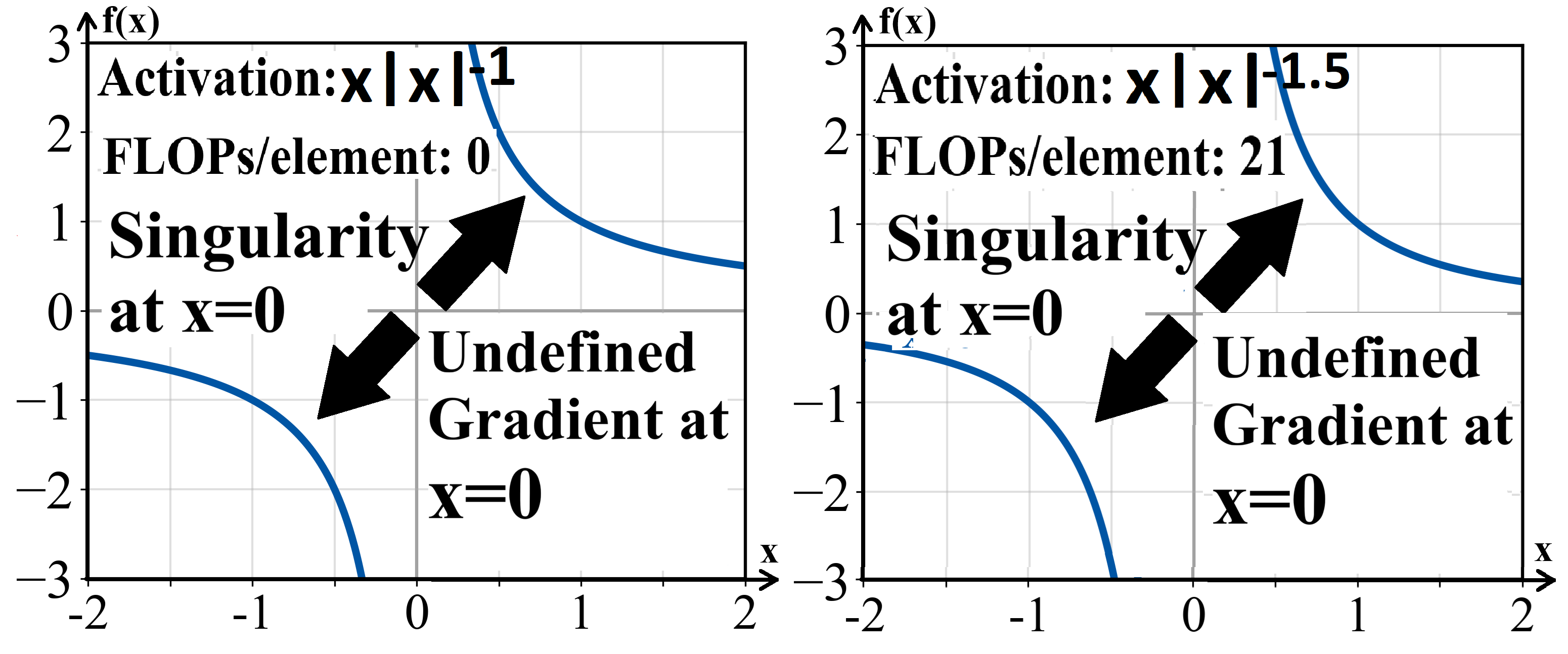}
     \vspace{-1.8em}
    \caption{{\color{red}LyRIC activation with $\alpha\in \{-1,-1.5\}$}.}
    \label{fig:LyRIC_discont}
    \vspace{-2.50em}
\end{figure}


The activation manifolds for different hyperparameter configurations exhibit \textbf{analogous topological characteristics} as shown in Fig. \ref{fig:LyRIC_main}. Nevertheless, configuring the expansion coefficient to $\alpha=1$ yields a considerably lower FLOPs footprint than $\alpha \in \{0.5,1.5,2,2.5,3\}$ \textit{(see Tables \ref{tab:flops_lyric} 
and \ref{tab:alphaLyRIC} for ablation on  $\alpha$)}. Therefore, we formalize our canonical LyRIC activation function as the following, ensuring well-behaved gradient flow while simultaneously achieving optimal  computational efficiency: 

\vspace{-0.4em}
\begin{equation}
    \small f(x) = x \vert x \vert^\alpha, \quad \text{where } \alpha=1
    \label{eqn6}
\end{equation}
 \vspace{-01.2em}

\noindent Fig.~\ref{fig:LyRIC_main} compares ReLU, GELU, and SiLU against LyRIC at $\alpha{=}1$ (left), and illustrates the effect of varying $\alpha \in \{0.5, 1.0, 1.5, 2.0, 2.5\}$ on the LyRIC activation shape (right). Furthermore, Table~\ref{tab:flops_lyric} shows that fractional values of $\alpha$ in the LyRIC activation ($P_0$, $P_2$, $P_4$) incur 
11--13 FLOPs/element due to non-integer exponentiation, while higher-order integers ($P_3$, $P_5$) reduce this to 2--3 FLOPs 
via successive multiplications. The canonical $\alpha{=}1$ ($P_1$) simplifies the leading term to a sign-preserving quadratic map, eliminating exponentiation and achieving the theoretical minimum of 1 FLOP/element while preserving LyRIC's topological properties. 
{\color{red}A rigorous analysis is provided in Appendix~\ref{subsec:Lyric_appendix}. Please note that our FLOPs/element calculations are based on an actual PyTorch implementation, 
which accounts for slight differences from standard mathematical estimations.}

\begin{table}[t]
\centering
\renewcommand{\arraystretch}{0.75} 
\resizebox{0.6\columnwidth}{!}{%
\begin{tabular}{l l c}
\toprule
\textbf{} & \textbf{Function} & \textbf{FLOPs/element$\downarrow$} \\
\midrule
$P_0$ & $\alpha=0.5$ & 11 \\
$P_1$ & $\alpha=1$ & 1 \\
$P_2$ & $\alpha=1.5$ & 12 \\
$P_3$ & $\alpha=2$ & 2 \\
$P_4$ & $\alpha=2.5$ & 13 \\
$P_5$ & $\alpha=3$ & 3 \\
\bottomrule
\end{tabular}}
 \vspace{-0.5em}
\caption{FLOPs footprint analysis of hyperparameter $\alpha$.}

\label{tab:flops_lyric}
\end{table}

  
\begin{figure}[t]
    \centering
    \includegraphics[width=1\columnwidth, height=5cm, keepaspectratio]{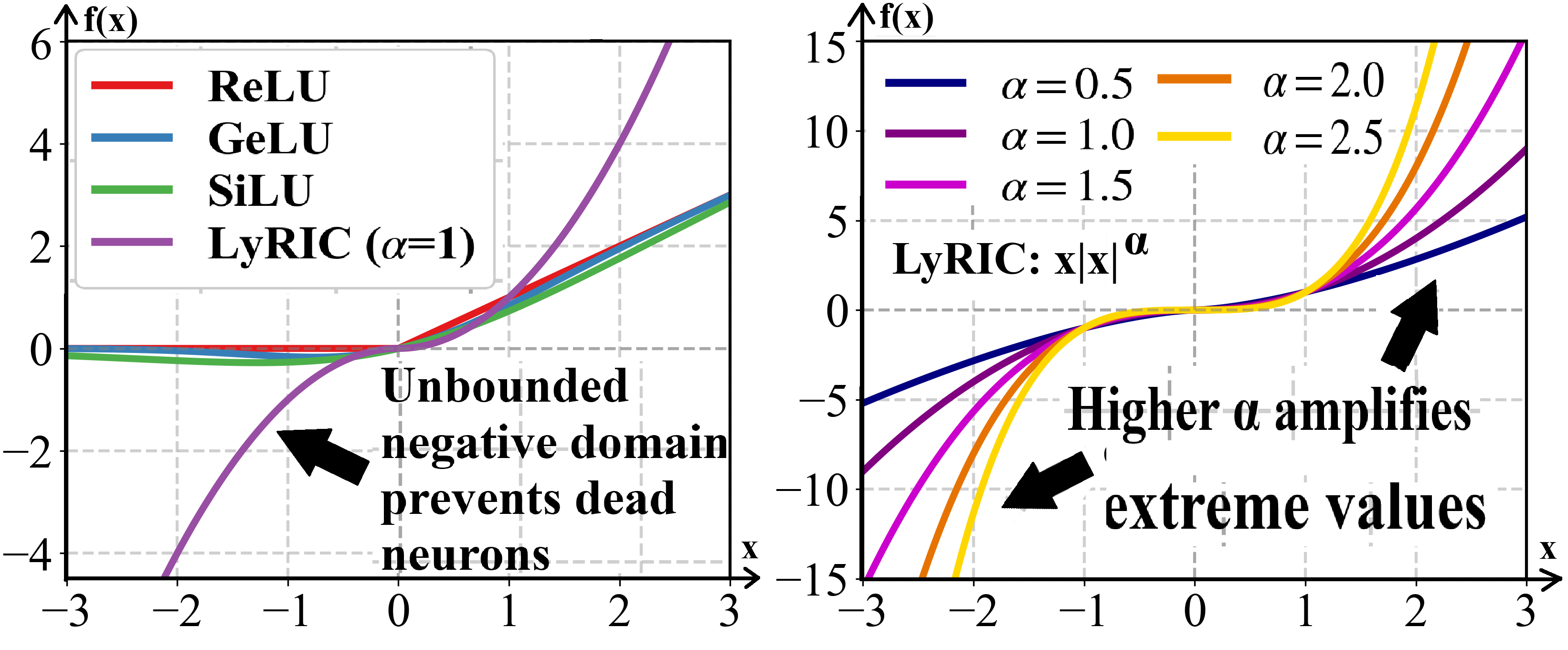}
    \caption{{\color{red}(Left) Comparison of different activation functions. (Right) Effect of $\alpha$ on LyRIC activation.}}
    \label{fig:LyRIC_main}
    \vspace{-0.50em}
\end{figure}

\subsection{Convolutional Feature Extraction (CFE)}
\label{subsec:CFE}

{\color{red}The CFE block (Fig. \ref{fig:cgs-ets}) comprises 3 residual blocks. Each contains two 1D convolutional layers with ReLU and a \textit{parameterized} skip connection ensuring dimensional congruency ({\color{black}Appendix \ref{subsec:CFE_appendix}}). A uniform channel capacity of $C_D=512$ is maintained across all layers. \textit{This design achieves superior EMG synthesis with a reduced parameter footprint ({\color{black}see Table \ref{tab:channel_dim} for channel ablation}).}}


\subsection{GLARE-fused Mamba Encoder--GLAME}
\label{subsec:flame}

Succeeding the CFE front-end, we introduce a novel encoder formulated to achieve strict parameter efficiency while robustly modeling the complex, non-linear temporal dependencies of  EMG signals. We empirically restrict the encoder depth to four layers (\textit{see  Fig. \ref{fig:cgs-ets} \& {\color{black}Table \ref{tab:Layer_Count}} for layer count ablation}). Two of the layers seamlessly integrate Mamba state-space block \cite{mamba2024}, our localized GLARE attention mechanism, and a terminal Feed Forward Network (FFN). We interleave the GLARE attention mechanism exclusively within the first and third layers of the encoder. This selective attention routing maximizes local context aggregation while sharply minimizing computational overhead (\textit{see {\color{black}Table \ref{tab:glare_attention}} for GLARE ablation}). Collectively, we name the four-layered encoder \textit{ (see Fig. \ref{fig:cgs-ets})} as GLAME (GLARE-fused Mamba Encoder). {\color{red}For multi-task learning, GLAME’s final latent representations bifurcate into a primary acoustic head mapping to 80-dimensional melspectrograms for waveform inversion, and an auxiliary linguistic head for phoneme-level supervision.}


\subsection{Design of the LER Loss}
\label{lerloss}

{\color{red}\textbf{Motivation:} We introduce a novel Lyapunov Exponent Regularization (LER) loss to capture the rapid, nonlinear fluctuations and sensitivity to initial conditions inherent in articulatory biosignals and dynamic properties that \textit{conventional reconstruction objectives typically overlook.} LER is built on Lyapunov Exponent (LE) \cite{oseledec1968multiplicative}, which quantifies the exponential rate of divergence between infinitesimally close state-space trajectories. \textit{\ul{As chaos is inherent in speech}, we use LE to reconstruct its underlying attractors of speech by directly penalizing mismatches in the local divergence rates.} This yields natural, temporally coherent acoustic features that preserve the complex micro-structure of human phonation.}

{
\setlength{\textfloatsep}{0pt plus 0pt minus 0pt}   
\begin{algorithm}[t]
\scriptsize
\caption{\small Logarithmic Divergence \& LER}
\label{alg:LER}
\begin{algorithmic}[1]
\REQUIRE EMG Feature sequence $x \in \mathbb{R}^{T \times C}$, $m=10$, $\tau=1$
\ENSURE Lyapunov exponent $\lambda$ and LER loss $\mathcal{L}_{\text{Lyap}}$
\STATE Compute $M \gets T - (m-1)\tau$, $d \gets m \cdot C$
\FOR{$i=0$ to $M-1$}
    \STATE $E_V[i] \gets [x[i,:], x[i+\tau,:], \dots, x[i+(m-1)\tau,:]]$
\ENDFOR
\FOR{$i,j \in \{0,\dots,M-1\}$}
    \STATE $D[i,j] \leftarrow \lVert E_V[i] - E_V[j] \rVert_2$
\ENDFOR
\FOR{$i=0$ to $M-1$}
    \STATE Find $j^*(i)$ minimizing $D[i,j]$ (exclude Theiler window)
    \FOR{$k=1$ to $k_{\max}$}
        \STATE $d_k[i] \leftarrow \lVert E_V[i+k] - E_V[j^*(i)+k] \rVert_2$
    \ENDFOR
\ENDFOR
\STATE $S(k) \gets \frac{1}{N_k} \sum_i \log(d_k[i]+\epsilon_1)$, Center $k$ and $S(k)$ to get $\tilde{k}, \tilde{S}$
\STATE Calculate $\lambda \gets \frac{\sum \tilde{k}_i \tilde{S}_i}{\sum \tilde{k}_i^2 + \epsilon_2}$ and  $\mathcal{L}_{\text{LER}} \gets |\lambda^{(\hat{Y}_S)} - \lambda^{(Y_V)}|$
\STATE \textbf{return} $\lambda$, $\mathcal{L}_{\text{LER}}$
\end{algorithmic}
\end{algorithm}
}

\textbf{Loss Calculation (\ul{see Alg.} \ref{alg:LER}):} The extracted aligned EMG features from the encoder output are passed to compute the logarithmic divergence $S(k)$ of nearest-neighbor pairs over time for each $k$, as described in Section~\ref{subsec:lyric} (\textbf{Row \textcircled{\scriptsize 14}}). The largest LE ($\lambda$) is then estimated as the weighted linear slope of $S(k)$: {\small $\lambda = \frac{\sum_{i=1}^{n_k} \tilde{k}_i \tilde{S}_i} {\sum_{i=1}^{n_k} \tilde{k}_i^2 + \epsilon_2}$} where $\epsilon_2 = 10^{-8}$ prevents division by zero. $\tilde{k}$ and $\tilde{S}$ represent the centered values of time offsets and logarithmic divergences, respectively. Then, we define LER loss as {\small $\mathcal{L}_{\text{LER}}(\hat{Y}_S[j], Y_V[i])$=$\big\vert \lambda^{(\hat{Y}_S[j])} - \lambda^{(Y_V[i])}\big\vert$}, where $\hat{Y}_S[j]$ and $Y_V[i]$ denote temporally aligned predicted and target audio features, respectively (\textbf{Row \textcircled{\scriptsize 15}}).

\subsection{Design of the MSDFA Loss}
\label{subsec:msdfaloss}

{\color{red}\textbf{Motivation:} We introduce a Multi-Scale Detrended Fluctuation Analysis (MSDFA) objective \cite{peng1994mosaic} to quantify the fractal-like, long-range temporal correlations overlooked by conventional frame-level spectrogram losses. \textit{Omitting the MSDFA objective causes the network to degenerate into generating smoothed, muffled prosody lacking natural dynamic variation, which underscores the critical necessity of long-range chaotic correlations for intelligible speech synthesis.}}

{
\setlength{\textfloatsep}{0pt plus 0pt minus 0pt}   
\begin{algorithm}[t]
\scriptsize
\caption{MSDFA Loss Computation}
\label{alg:MSDFA}
\begin{algorithmic}[1]
\REQUIRE EMG Feature sequence $x \in \mathbb{R}^{T \times C}$, scales $s$
\FOR{$c=1$ to $C$}
    \STATE $\mu_c \leftarrow \text{mean}(x[:,c])$
    \STATE $Y_s \leftarrow \text{cumsum}(x[:,c]-\mu_c)$
    \STATE Construct $X_s$ with rows $[1,\tau_j]$, $\tau_j=(j-1)/(s-1)$
    \STATE $P_s \leftarrow X_s(X_s^T X_s)^{-1}X_s^T$, \quad $K_s \leftarrow I_s-P_s$
    \STATE $R_s \leftarrow Y_s K_s$  \text (detrended residuals)
    \STATE Compute $F[c,i]$ as RMS over $R_s$ for each scale
    \STATE $\alpha_c \leftarrow \frac{\sum \log s[i]\log F[c,i]}{\sum \log s[i]}^2+\epsilon_3$
\ENDFOR
\STATE $\mathcal{L}_{\text{MSDFA}} \leftarrow \frac{1}{|\mathcal{V}|}\sum_{c\in\mathcal{V}} |\alpha_c^{(\hat{Y}_S)}-\alpha_c^{(Y_V)}|$
\end{algorithmic}
\end{algorithm}
}

\textbf{Loss Calculation (\ul{see Alg.} \ref{alg:MSDFA}):} For each channel $c \in \{1, \dots,C\}$, we compute the mean $\mu_c$ and define the profile (integrated features) as the cumulative sum of deviations from the mean $y[k,c]$ where $k \in \{1,\dots,T\}$ and $T$ is the number of time frames. This sum constructs the windowed profile tensor $Y_s$ (\textbf{Rows \textcircled{\scriptsize 1}--\textcircled{\scriptsize 3}}). The design matrix $X_s \in \mathbb{R}^{s \times 2}$ is defined with rows $[1, \tau_j]$ where $\tau_j$=$(j-1)/(s-1)$ represents the normalized time within the segment for $j \in \{1, \dots, s\}$ and the projection matrix onto the column space of $X_s$ is $P_s$=$X_s(X_s^T X_s)^{-1} X_s^T \in \mathbb{R}^{s \times s}$. The detrending kernel can be denoted by $K_s$=$I_s - P_s \in \mathbb{R}^{s \times s}$ where $I_s \in \mathbb{R}^{s \times s}$ is the identity matrix. The detrended residuals are computed as $R_s[c, i, j]$=$\sum_{j'=1}^s Y_s[c, i, j'] \cdot K_s[j', j]$ (\textbf{Rows \textcircled{\scriptsize 4}--\textcircled{\scriptsize 7}}). Here, $i \in \{1, \dots, N_s\}$ and $j \in \{1, \dots, s\}$ are segment (window) index and position within the segment, respectively. The MSDFA scaling component for channel $c$ is calculated using the \textbf{Row \textcircled{\scriptsize 8}}, where $\epsilon_3$=$10^{-8}$ prevents division by zero. The $\log s[i]$ and $\log F[c,i]$ are centered logarithmic values, which are root-mean-square across all $R_s$. The $s$ and $F$ are pre-defined set of scales for MSDFA function. Given, temporally aligned $\hat{Y}_S[j]$ and $Y_V[i]$, we define MSDFA loss as $\mathcal{L}_{\text{MSDFA}}$=$\frac{1}{|\mathcal{V}|} \sum_{c \in \mathcal{V}} \big\vert \alpha_c^{(\hat{Y}_S[j])}$-$\alpha_c^{(Y_V[i])} \big\vert$, where $|\mathcal{V}|$ is cardinality of the valid channel set.

\subsection{Total Loss Function}

We employ Dynamic Time Warping (DTW), Pairwise Distance (PD) loss ($\mathcal{L}_{PD}$), and phoneme loss ($\mathcal{L}_{Ph}$) from \cite{gaddy-klein-2021-improved}, together with our two novel \textit{chaos-inspired} loss functions, $\mathcal{L}_{LER}$ and $\mathcal{L}_{MSDFA}$. The  $\mathcal{L}_{LER}$ is designed to capture chaotic temporal irregularities, whereas $\mathcal{L}_{MSDFA}$ models chaotic multi-scale dynamics. {\color{red}The encoder outputs are temporally aligned at the frame level using DTW.} These aligned bins are then used to compute the LE and MSDFA. The total loss $\mathcal{L}_{\text{total}}$:

\vspace{-1.9em}
\begin{equation}
\mathcal{L}_{total} = \mathcal{L}_{PD}+\mathcal{L}_{Ph}+\mathcal{L}_{LER}+\mathcal{L}_{MSDFA}   
\label{eqn:totalloss}
\end{equation}
    
\section{Comprehensive Analysis}
\label{sec:Comprehensive_Result_Analysis}

\subsection{Evaluation Metrics}
\label{subsec:EvaluationMetrics}

{\color{red}After applying post-vocoder alignment \textit{(see Appendix \ref{subsec:inference_workflow} and Fig. \ref{fig:cgs-ets} for details)}, we use {ten} evaluation} metrics -- Log Spectral Distance (LSD), Short-Time Objective Intelligibility (STOI), Perceptual Evaluation of Speech Quality (PESQ), Scale-Invariant Signal-to-Distortion Ratio (SI-SDR), Non-Intrusive Speech Quality Assessment Mean Opinion Score (NISQA-MOS), Floating-Point Operations (FLOPs), Inference Time (IT) per batch, Word Error Rate (WER), {\color{red}Character Error Rate (CER)} and Real-Time Factor (RTF). \textit{No prior work in the literature has employed such an extensive set of evaluation metrics for ETS systems.}

\subsection{Dataset and Hyperparameter} 
\label{subsec:HyperparameterConfiguration}


We evaluate on the state-of-the-art (SOTA) public English dataset \cite{gaddy-klein-2020-digital}, comprising 19 hours of facial EMG data collected under silent and vocalized speech conditions. {\color{black}We do minimal preprocessing to facilitate stable model optimization.} {\color{red}We also evaluate on a 6-channel \textit{multispeaker} Mandarin dataset AVE Speech \cite{11079657}, consisting of $\sim$55 hours of synchronized audio and EMG data from 100 speakers \textit{(see Appendix \ref{sec:dataset_preprocessing} for detailed data preparation and preprocessing on both datasets)}. {We ablate on unmodified EMG data, which already contains inherent real-world noise (cross-talk, shifts, interference) to validate robustness under noisy conditions.}}


{\color{black} We train for 80 epochs with batch size 32 using Adaptive Moment Estimation with Weight Decay (AdamW) optimizer having a learning rate, $\eta=10^{-3}$ and weight  decay $=10^{-7}$. Gradients are clipped to norm 0.5. {\color{red}Models are trained in FP32 precision using distributed data parallelism across two NVIDIA RTX 4090 GPUs} \textit{(see Appendix~\ref{subsec:hyperparameters} for hyperparameter and configuration details)}.
}

\subsection{Comparative Analysis with Baselines} 
\label{subsec:comparison}


\begin{table}[t]
\color{red}
    \centering
    
    \setlength{\tabcolsep}{2.1pt}     
    \renewcommand{\arraystretch}{1.0} 
    
    \resizebox{\columnwidth}{!}{%
    \begin{tabular}{ l | l | l | c | c | c | c | c | c | c | c }
    \toprule
    & \textbf{Model} & \textbf{Cr} & \textbf{L$\downarrow$} & \textbf{ST$\uparrow$} & \textbf{P$\uparrow$} & \textbf{SD$\uparrow$} & \textbf{N$\uparrow$} & \textbf{W$^*\downarrow$} & \textbf{C$\downarrow$} & \textbf{Pr$\downarrow$} \\
    \midrule
    
    $P_0$ & Baseline$^{1}$ & EN & 3.23 & 0.04 & 1.17 & -52.5 & 3.28 & 68.00 & 38.2 & 60.21 \\
    $P_1$ & Baseline$^{1}$ & ZH & 3.50 & 0.03 & 1.05 & -58.1 & 1.31 & $\times$ & 59.1 & 60.21 \\
     \hline
    
    $P_2$ & Baseline$^{2}$ & EN & 2.25 & 0.13 & 1.18 & -42.0 & 3.30 & 42.20 & 23.7 & 54.10 \\
    $P_3$ & Baseline$^{2}$ & ZH & 2.67 & 0.10 & 1.08 & -39.5 & 1.84 & $\times$ & 39.8 & 54.10 \\
    \hline
    
    $P_4$ & Baseline$^{3}$ & EN & $\times$ & $\times$ & $\times$ & $\times$ & $\times$ & 46.14 & $\times$ & $\times$ \\
   \hline
    $P_5$ & Baseline$^{4}$ & ZH & $\times$ & $\times$ & $\times$ & $\times$ & $\times$ & $\times$ & 38.0 & $\times$ \\
    \hline
    \hline
    
    {$P_6$} & {CGS-ETS} & {EN} & {1.08} & {0.61} & {1.19} & {-32.6} & {3.33} & {41.36} & {23.2} & {14.34} \\
    {$P_7$} & {CGS-ETS} & {ZH} & {1.11} & {0.40} & {1.09} & {-34.41} & {2.20} & {$\times$} & {36.3} & {14.34} \\
    \bottomrule
    \multicolumn{11}{@{}l@{}}{Baseline$^{1}$\cite{gaddy-klein-2020-digital} \quad $^{2}$\cite{gaddy-klein-2021-improved}} \\
    \multicolumn{11}{@{}l@{}}{$^{3}$\cite{scheck2023} \quad $^{4}$\cite{xie2025neural}} \\
    \end{tabular}%
    } 
    \vspace{-0.5em}
.    \caption{\normalfont \color{red}Comparison against SOTA methods. Here, Cr=Corpus, EN=English, ZH=Mandarin, L=LSD, ST=STOI, P=PESQ, SD=SI-SDR, N=NISQA-MOS, W=WER \%, C=CER \%, Pr=Parameters in Millions. $^*$As Mandarin text lacks spaces, corpora rely on CER rather than WER. $P_4$ and $P_5$ codebases are not public.}
    \label{tab:comparison}
     \vspace{-0.8em}
\end{table}
{\color{red}
Table \ref{tab:comparison} comprehensively evaluates our CGS-ETS against four recent SOTA baselines}: \cite{gaddy-klein-2020-digital, gaddy-klein-2021-improved, scheck2023, xie2025neural} on {\color{red}English and Mandarin datasets. Historically, most prior works report only text-based error rates like WER or CER, and none of them do point-to-point quality assessment due to temporal misalignment issues in the post-vocoder stage. A full suite of perceptual and structural metrics (LSD, STOI, PESQ, SI-SDR) requires point-to-point (frame-by-frame) speech calculation. To solve this issue, we use DTW at the post-vocoder stage to ensure sample-by-sample comparison.

Comparing against the strongest fully evaluated baseline$^2$ ($P_2$) on the English corpus reveals that CGS-ETS establishes a definitive new standard for acoustic reconstruction and fidelity. Our chaos-inspired framework with the GLARE attention delivers a \textbf{2.08$\times$ improvement} in spectral reconstruction accuracy (LSD = 1.08 vs 2.25), a \textbf{4.69$\times$ increase} in objective speech intelligibility (STOI = 0.61 vs 0.13), and a \textbf{1.28$\times$ enhancement} in noise suppression (SI-SDR = -32.6 vs -42.0). {\color{red}CGS-ETS yields an improved WER of 41.36\% and a CER of 23.2\%.} \textit{Please note that CGS-ETS achieves reconstruction improvement with a significant degree of architectural compression.} As detailed in Table \ref{tab:comparison}, our model operates with a mere \textbf{14.34M parameters}, contrasting with the highly parameter-dense \textbf{54.10M} requirement of baseline$^2$ ($P_2$, $P_3$). This constitutes a remarkable \textbf{73.49\% reduction} in total model capacity. {Integrating non-linear chaotic dynamics and bounding local context (GLARE) provides a superior but structurally compact inductive bias for ETS synthesis than relying on the brute-force mapping of massive multi-head attention transformers.} Please note that we use baselines ($P_0$--$P_3$) for full point-to-point comparisons, as recent SOTA works \cite{scheck2023, xie2025neural} ($P_4$, $P_5$) do not report model size or computational complexity, and their codebases are not publicly available, precluding direct efficiency comparisons. {\color{red}However, \cite{ren2024} calculates an LSD of 1.86 in their recent work. In comparison, our model achieves an LSD of 1.08 on the English corpus, demonstrating a substantial 41.9\% improvement. Furthermore, the recent Mandarin baseline \cite{xie2025neural} reports a CER of 38.0\%. Our CGS-ETS ($P_7$) achieves a superior CER of 36.3\% and surpasses all other baselines on all other metrics, demonstrating a definitive improvement on the multispeaker Mandarin corpus as well.}}

\begin{table}[t]
    \centering
    \color{red}
    
    \setlength{\tabcolsep}{1.5pt} 
    \renewcommand{\arraystretch}{1.0}
    \resizebox{\columnwidth}{!}{%
    \begin{tabular}{@{} l | c | c | c | c | c | c @{}}
    \toprule
    \textbf{Model} & \textbf{CFE$\downarrow$} & \textbf{Enc$\downarrow$} & \textbf{Total$\downarrow$} & \textbf{IT$\downarrow$} & \textbf{Fs$\downarrow$} & \textbf{RTF$\downarrow$} \\
    \midrule
    \cite{gaddy-klein-2021-improved} & 10.17 & 44.03 & 54.10 & 5.97 & 2400 & 0.0057 \\
    {Proposed CGS-ETS} & {4.49} & {9.85} & {14.34} & {5.61} & {926.6} & {0.0057} \\
    \bottomrule
    \end{tabular}%
    } 
    \caption{\normalfont \color{red}Computational efficiency. Here, Enc=Encoder Size, Total = Total Parameters in Millions, ITs = Inference Time in Milliseconds, and Fs = FLOPs in Millions.}
    \label{tab:computational_complexity}
    \vspace{-0.98em}
\end{table}

\subsection{Computational Complexity}
\label{subsec:computational_complexity}

{\color{red} Table~\ref{tab:computational_complexity} compares CGS-ETS's complexity against the best-performing baseline \cite{gaddy-klein-2021-improved} on the English dataset. CGS-ETS achieves massive parameter reductions across all components, slashing the Encoder size by \textbf{77.6\%} (44.03M$\to$9.85M) and the CFE by \textbf{55.8\%} (10.17M$\to$4.49M), yielding a \textbf{73.5\%} decrease in total size (54.10M$\to$14.34M). FLOPs drop by \textbf{61.4\%} (2400M$\to$926.6M) and inference time improves to 5.61ms. CGS-ETS maintains real-time capabilities (RTF: 0.0057) while simultaneously improving audio quality \textit{(see Table~\ref{tab:comparison})}.}

\subsection{Ablation Study} 
\label{Ablation_Study}

{\color{red}Here, all ablation studies are conducted using the English dataset from \cite{gaddy-klein-2020-digital}.}

{\color{red}\subsubsection{Chaos Loss Components}}
\label{subsubsec:loss_ablation}

{\color{red}Table \ref{tab:loss_ablation} isolates the contributions of our novel chaotic losses to CGS-ETS. Isolated LER ($P_2$) provides localized gradient regularization, dropping WER by \textbf{5.85\%} (51.56\% $\to$ 48.34\%). Meanwhile, MSDFA ($P_3$) captures macro-level fractal correlations, reducing WER by \textbf{11.47\%} (51.56\%$\to$45.65\%) and boosting naturalness (NISQA: 3.31). Optimizing both ($P_4$) mathematically constrains the \textit{deterministic chaotic speech invariants}, achieving optimal results: minimizing WER to \textbf{41.36\%} while maximizing acoustic fidelity (STOI: 0.61, LSD: 1.08, SI-SDR: -32.61).}


\begin{table}[t]
    \centering
    \color{red}
    \footnotesize 
    
    \setlength{\tabcolsep}{1.5pt}     
    \renewcommand{\arraystretch}{1.0}
    \resizebox{\columnwidth}{!}{%
    \begin{tabular}{@{} l | l | c | c | c | c | c | c | c @{}}
    \toprule
    & \textbf{Model Config.} & \textbf{L$\downarrow$} & \textbf{ST$\uparrow$} & \textbf{P$\uparrow$} & \textbf{SD$\uparrow$} & \textbf{N$\uparrow$} & \textbf{W$\downarrow$} & \textbf{Pr$\downarrow$} \\
    \midrule
    
    
    $P_1$ & w/o LER+MSDFA & 1.14 & 0.60 & 1.12 & -35.5 & 3.25 & 51.56 & 14.34 \\
    $P_2$ & w LER w/o MSDFA & 1.10 & 0.60 & 1.18 & -33.7 & 3.29 & 48.34 & 14.34 \\
    $P_3$ & w MSDFA w/o LER & 1.09 & 0.60 & 1.19 & -32.6 & 3.31 & 45.65 & 14.34 \\
    \midrule
    {$P_4$} & {w LER+MSDFA} & {1.08} & {0.61} & {1.19} & {-32.6} & {3.33} & {41.36} & {14.34} \\
    \bottomrule
    
    \end{tabular}%
    } 
    \vspace{-0.5em}
     \caption{\normalfont \color{red} Ablation study on chaotic losses. Here, metrics acronyms are provided in Tables \ref{tab:comparison} and \ref{tab:computational_complexity}.}
    \label{tab:loss_ablation}
\end{table}

\subsubsection{Activation Functions}
\label{subsubsec:activation_function}

{\color{red}Table~\ref{tab:activation_function} evaluates CFE, Mamba, and FFN activations (stable LSD, STOI, and PESQ omitted; see Appendix \ref{appendix:activation_function}). Swapping SwiGLU for GeLU in the FFN ($P_0 \to P_1$) cuts FLOPs and WER (51.16\%$\to$47.06\%). LyRIC in CFE ($P_2$) or FFN ($P_3$) destabilizes acoustics. Anchoring LyRIC solely in Mamba ($P_4$) yields peak spectral scores, but pairing with SwiGLU inflates computation. The optimal setup ($P_5$) LyRIC-driven Mamba with GeLU achieves best WER (41.36\%), NISQA (3.33), and minimal inference cost (RTF: 0.0057).}


\begin{table}[t]
    \centering
    \color{red}
    \Large 
    
    \setlength{\tabcolsep}{1.5pt}     
    \renewcommand{\arraystretch}{1.0}
    
    \resizebox{\columnwidth}{!}{%
    \begin{tabular}{@{} l | l | l | l | c | c | c | c | c | c @{}}
    \toprule
    & \textbf{CFE} & \textbf{Mam.} & \textbf{FFN} & \textbf{SD$\uparrow$} & \textbf{N$\uparrow$} & \textbf{W$\downarrow$} & \textbf{IT$\downarrow$} & \textbf{Fs$\downarrow$} & \textbf{RTF$\downarrow$} \\
    \midrule
    
    $P_0$ & ReLU & SiLU & SwiGLU & -31.5 & 3.25 & 51.16 & 7.12 & 1038 & 0.0075 \\
    $P_1$ & ReLU & SiLU & GeLU   & -31.3 & 3.28 & 47.06 & 5.92 & 929.8 & 0.0059 \\
    $P_2$ & LyRIC& SiLU & GeLU   & -33.2 & 2.90 & 71.86 & 5.92 & 929.8 & 0.0059 \\
    $P_3$ & ReLU & SiLU & LyRIC  & -31.9 & 3.23 & 52.63 & 5.70 & 928.4 & 0.0057 \\
    $P_4$ & ReLU & LyRIC& SwiGLU & -32.6 & 3.30 & 45.71 & 6.21 & 1035 & 0.0062 \\
    \midrule
    
    {$P_5$} & {ReLU} & {LyRIC} & {GeLU} & {-32.6} & {3.33} & {41.36} & {5.61} & {926.6} &{0.0057} \\
    \bottomrule
    \end{tabular}%
    } 
    \caption{\normalfont \color{red}Ablation on activations. Here, metrics acronyms are provided in Tables \ref{tab:comparison} and \ref{tab:computational_complexity}.}
    \label{tab:activation_function}
     \vspace{-0.7em}
\end{table}

\subsubsection{GLARE Attention Depth}
\label{subsubsec:glare_attention}

{\color{red}Table~\ref{tab:glare_attention} ablates GLARE depth routing (stable metrics omitted; see Appendix \ref{appendix:glare_attention}). Removing attention ($P_0$) spikes WER to 73.22\%, while saturating all layers ($P_6$) inflates computation and redundancy (WER: 44.56\%). Imbalanced placements ($P_1$--$P_5$) stall WER at 45--50\%. The alternating setup ($P_7$, Layers 1 \& 3) optimally balances local attention and global mappings, achieving the best WER (41.36\%) and NISQA (3.33) with minimal overhead (14.34M Params; IT: 5.61ms).}


\begin{table}[t]
    \centering
    \color{red}
    \Large 
    
    \setlength{\tabcolsep}{1.5pt}     
    \renewcommand{\arraystretch}{1.0} 
    
    \resizebox{\columnwidth}{!}{%
    \begin{tabular}{@{} l | l | c | c | c | c | c | c | c @{}}
    \toprule
    & \textbf{Layers} & \textbf{SD$\uparrow$} & \textbf{N$\uparrow$} & \textbf{W$\downarrow$} & \textbf{Pr$\downarrow$} & \textbf{IT$\downarrow$} & \textbf{Fs$\downarrow$} & \textbf{RTF$\downarrow$} \\
    \midrule
    $P_0$ & None        & -34.4 & 2.91 & 73.22 & {12.68} & {4.23} & {848.3} & {0.0046} \\
    $P_1$ & $L=1,2$     & -32.4 & 3.30 & 46.14 & 14.34 & 5.91 & 926.6 & 0.0059 \\
    $P_2$ & $L=3,4$     & -33.0 & 3.25 & 50.25 & 14.34 & 5.97 & 926.6 & 0.0059 \\
    $P_3$ & $L=1,4$     & -33.5 & 3.30 & 49.01 & 14.34 & 5.91 & 926.6 & 0.0061 \\
    $P_4$ & $L=1,2,3$   & -33.3 & 3.26 & 46.38 & 15.05 & 6.86 & 966.5 & 0.0068 \\
    $P_5$ & $L=2,3,4$   & -32.8 & 3.28 & 44.91 & 15.05 & 7.11 & 966.5 & 0.0071 \\
    $P_6$ & $L=1,2,3,4$ & -31.9 & 3.24 & 44.56 & 15.83 & 7.58 & 1006 & 0.0077 \\
    \midrule
    {$P_7$} & {$L=1,3$} & {-32.6} & {3.33} & {41.36} & {14.34} & {5.61} & {926.6} & {0.0057} \\
    \bottomrule
    \end{tabular}%
    } 
    \caption{\normalfont \color{red} Ablation study on GLARE attention on different layer depths. Metrics LSD, STOI, and PESQ showed negligible variance, hence are omitted here. Metrics acronyms are provided in Tables \ref{tab:comparison} and \ref{tab:computational_complexity}.}
    \label{tab:glare_attention}
\end{table}

\subsubsection{Hyperparameter $\alpha$ in LyRIC Activation}
\label{subsubsec:alphaLyRIC}

{\color{red}Table~\ref{tab:alphaLyRIC} ablates the LyRIC expansion coefficient $\alpha$ (stable LSD, STOI, and PESQ omitted). Setting $\alpha=0$ ($P_0$) yields a linear identity, degrading WER to 48.22\%. Fractional $\alpha=1.5$ ($P_1$) incurs exponentiation overhead. Higher orders ($\alpha \in \{2,3\}$; $P_2, P_3$) destabilize gradients, spiking WER to 49.26\%. The canonical quadratic $\alpha=1$ ($P_4$) achieves the optimal trade-off: lowest WER (41.36\%), highest NISQA (3.33), and minimal inference cost (926.6M FLOPs; RTF: 0.0057).}


\begin{table}[t]
\centering
\color{red}
\setlength{\tabcolsep}{1.5pt} 
\small
\renewcommand{\arraystretch}{1.0}
\resizebox{\columnwidth}{!}{%
\begin{tabular}{ l | l | c | c | c | c | c | c | c | c | c}
\toprule
\textbf{} & \textbf{$\alpha$} &
\textbf{L$\downarrow$} & \textbf{ST$\uparrow$} & \textbf{P$\uparrow$} &
\textbf{SD$\uparrow$} & \textbf{N$\uparrow$} & \textbf{W$\downarrow$} &
\textbf{IT$\downarrow$} & \textbf{Fs$\downarrow$} & \textbf{RTF$\downarrow$} \\
\midrule
$P_0$ & $0$   & 1.09 & 0.60 & 1.18 & -34.1 & 3.30 & 48.22 & 5.98 & 927.0 & 0.0058 \\
$P_1$ & $1.5$ & 1.09 & 0.61 & 1.18 & -30.8 & 3.26 & 44.73 & 5.95 & 929.9 & 0.0060 \\
$P_2$ & $2$   & 1.08 & 0.60 & 1.18 & -32.3 & 3.30 & 45.40 & 5.95 & 927.2 & 0.0059 \\
$P_3$ & $3$   & 1.09 & 0.60 & 1.18 & -33.0 & 3.30 & 49.26 & 6.00 & 927.3 & 0.0059 \\
\midrule
{$P_4$} & {$1$} & {1.08} & {0.61} & {1.19} & {-32.6} & {3.33} & {41.36} & {5.61} & {926.6} & {0.0057} \\
\bottomrule
\end{tabular}}
\caption{\color{red} Ablation study on hyperparameter $\alpha$ of LyRIC activation. Acronyms are provided in Tables \ref{tab:comparison} and \ref{tab:computational_complexity}.}
\vspace{-0.9em}
\label{tab:alphaLyRIC}
\end{table}


\subsubsection{Different Attention Mechanisms}
\label{subsubsec:Swa_ablation}
\vspace{-0.2em}

\begin{table}[ht!]
\centering
\color{red}
\Large
 
\setlength{\tabcolsep}{1.5pt} 
\renewcommand{\arraystretch}{1.0}
\resizebox{\columnwidth}{!}{%
\begin{tabular}{l | l | c | c | c | c | c | c | c}
\toprule
\textbf{} & \textbf{Attention Type} & \textbf{SD$\uparrow$} & \textbf{N$\uparrow$} & \textbf{W$\downarrow$} & \textbf{Enc$\downarrow$} & \textbf{IT$\downarrow$} & \textbf{Fs$\downarrow$} & \textbf{RTF$\downarrow$} \\
\midrule
$P_0$ & MHA, All Layers & -32.4 & 3.30 & 43.32 & 14.49 & 7.42 & 1166 & 0.0074 \\
$P_1$ & MHA, $L=1,3$ & -32.4 & 3.27 & 44.73 & 12.39 & 5.92 & 1061 & 0.0059 \\
$P_2$ & MQA, All Layers & -31.7 & 3.33 & 44.85 & 10.56 & 7.10 & 966.5 & 0.0071 \\
$P_3$ & MQA, $L=1,3$ & -31.8 & 3.31 & 44.90 & 9.38 & 5.69 & 907.4 & 0.0058 \\
\midrule
{$P_4$} & {GLARE, $L=1,3$} & {-32.6} & {3.33} & {41.36} & {9.85} & {5.61} & {926.6} & {0.0057} \\
\bottomrule
\end{tabular}}
\caption{\color{red} Ablation study on different attention mechanisms. Here, Enc=Encoder Size in Millions. Other metrics acronyms are provided in Tables \ref{tab:comparison} and \ref{tab:computational_complexity}.}
\label{tab:Swa_ablation}
\end{table}

{\color{red}\noindent Table~\ref{tab:Swa_ablation} ablates attention mechanisms (stable metrics omitted; see Appendix \ref{appendix:swa_ablation}). All-layer MHA ($P_0$) achieves 43.32\% WER but demands 1166M FLOPs. Truncating MHA ($P_1$) or using MQA ($P_2, P_3$) cuts FLOPs but degrades WER to ~45\%. In contrast, our proposed GLARE ($P_4$) yields the optimal trade-off: achieving the lowest WER (41.36\%) and highest SI-SDR (-32.6) while maintaining faster inference (5.61ms; 926.6M FLOPs). These support the fact that GLARE can be an efficient alternative to standard MHA for ETS tasks.}


\vspace{-0.2em}
\subsection{\color{red}Statistical Analysis of MOS Scores}
\vspace{-0.0em}

{\color{red}We evaluated {CGS-ETS} against the \cite{gaddy-klein-2021-improved} baseline using NISQA-MOS (Tables \ref{tab:mos_descriptive} and \ref{tab:wilcoxon_results}). At the utterance level (number of samples, $n=99$), our model achieved a higher mean NISQA-MOS (3.329 vs. 3.301, $p = 0.807$). {CGS-ETS} shows better stability. At the session level ($n=7$), it substantially reduces variance, yielding a Standard Deviation (SDv) of 0.059 and Standard Error of the Mean (SEM) of 0.022, compared to the baseline's 0.195 and 0.074 \textit{(see Appendix~\ref{sec:comprehensive_stats} for details)}. This tighter spread indicates our architecture is considerably more robust to non-stationary EMG across recording sessions, which is a critical requirement for reliable real-world SSIs.
}

\begin{table}[t]
\centering
\color{red}
\small
\setlength{\tabcolsep}{1.5pt} 
\renewcommand{\arraystretch}{1.0}
\resizebox{\columnwidth}{!}{%
\begin{tabular}{l | l | c | c | c | c | c}
\toprule
\textbf{Agg.} & \textbf{Model} & $n$ & \textbf{Mean} $\uparrow$ & \textbf{SDv} & \textbf{SEM} & \textbf{95\% CI} $\uparrow$\\
\midrule
Utter. & Gaddy 2021 & 99 & 3.301 & 0.593 & 0.060 & [3.193, 3.409] \\
Utter. & CGS-ETS   & 99 & 3.329 & 0.560 & 0.056 & [3.217, 3.440] \\
\midrule
Sess. & Gaddy 2021 & 7  & 3.303 & 0.195 & 0.074 & [3.132, 3.474] \\
Sess. & CGS-ETS   & 7  & 3.328 & 0.059 & 0.022 & [3.273, 3.383] \\
\bottomrule
\end{tabular}
}
\caption{\color{red}MOS (NISQA) descriptive statistics.}
\label{tab:mos_descriptive}
 \vspace{-0.8em}
\end{table}

\begin{table}[t]
\centering
\color{red}
\setlength{\tabcolsep}{1.5pt} 
\renewcommand{\arraystretch}{1.0}
\resizebox{\columnwidth}{!}{%
\begin{tabular}{l | c | c | c | c | c | c | c}
\toprule
\textbf{Agg.} & $n$ & $\overline{\textbf{d}}$ & $\textbf{Median(d)}$ & {$W^+$} & $W^-$ & $\textbf{T}$ & $\textbf{p}$ \\
\midrule
Utter. & 99 & 0.028 & -0.064 & 2405.0 & 2545.0 & 2405.0 & 0.807 \\
Sess.  & 7  & 0.025 & -0.008 & 13.0   & 15.0   & 13.0   & 0.938 \\
\bottomrule
\end{tabular}
}
\caption{\color{red}Paired Wilcoxon signed-rank test results.}
\label{tab:wilcoxon_results}
\vspace{-0.9em}
\end{table}

\subsection{\color{red}Additional Ablation Study}
\label{subsubsec:additional_ablation}

{\color{red}
Appendix \ref{subsubsec:channel_dim} and \ref{subsec:Layer_Count} show how channel capacity and the hierarchical depth of GLAME impact the model's performance.} {\color{red}Appendix \ref{appendix:disentanglement} shows the summary of the individual contributions of GLARE, GLAME, LyRIC, LER, and MSDFA.}

\subsection{\color{black}Subjective Mean Opinion Score (MOS)}
\label{subsec:Subjective Test}

{\color{red}
We use 5-point (1=bad to 5=excellent) MOS ratings for subjective evaluation. The unprocessed EMG has MOS=1, meaning completely unrecognizable. \textit{Our CGS-ETS (MOS=4.25), outperforms \cite{gaddy-klein-2021-improved} (MOS=3.98).} 
We refer to Appendix \ref{subsec:subj_eval_details} for details on MOS analysis.
}

\section{Conclusion} 
{\color{red}
CGS-ETS improves ETS architectures using chaos theory. Integrating \textit{\textbf{LyRIC}} and \textit{\textbf{GLARE}} within GLAME, alongside chaotic objectives {\color{red}\textit{(\textbf{LER} and \textbf{MSDFA})}, establishes strong theoretical grounding and introduces detailed evaluation metrics.} On the Gaddy benchmark~\cite{gaddy-klein-2021-improved}, it achieves superior fidelity and intelligibility with 73.49\% fewer parameters, validating chaotic dynamics as an efficient prior for real-time SSIs.
}}


\section{Acknowledgment}  
{\color{red}We are thankful to Gaddy et al. \cite{gaddy-klein-2020-digital} and Zhou et al. \cite{11079657} for releasing the ETS datasets as open-source.} We also express gratitude to the participants of the subjective evaluation tests for their contributions to the listening tests. We acknowledge that we have used ChatGPT for debugging codes, and finding grammatical errors to improve the manuscript.  

\section{Limitations}
{\color{red} Our current work establishes a new baseline using the single-speaker English dataset from \cite{gaddy-klein-2020-digital}, which serves as the gold standard for EMG-to-speech synthesis, as well as a multi-speaker Mandarin dataset from \cite{11079657}. However, these datasets do not consider 
multi-modal environments, such as visual images or video. Therefore, we focus our design to robustly handle intrinsic physiological artifacts for speech synthesis. }

{\color{red} CGS-ETS introduces specific computational constraints during the training phase while it achieves state-of-the-art performance in multispeaker EMG-to-Speech synthesis. The integration of the LER and MSDFA requires the reconstruction of high-dimensional phase-space manifolds. For an utterance of $T=200$ frames with $C=80$ features, the LER estimation involves computing a distance matrix across a concatenated feature vector of $m \times C$ dimensions. As shown in Table \ref{tab:overhead_complexity_analysis}, this introduces a training-time overhead where the LER calculation consumes approximately $31.8$ M MACs per segment, whereas the MSDFA overhead is computationally minimal ($0.358$ M MACs) due to the utilization of pre-computed detrending kernels. 

\begin{table}[ht]
\centering
\color{red}
\Large
\scriptsize
\setlength{\tabcolsep}{1.5pt} 
\renewcommand{\arraystretch}{1.0}
\resizebox{\columnwidth}{!}{
\begin{tabular}{l c | c | c | c | c | c}
\toprule
   & \multicolumn{3}{c}{\textbf{Lyapunov (LER)}} & \multicolumn{3}{c}{\textbf{Fractal (MSDFA)}}\\
\cmidrule(lr){2-4}\cmidrule(lr){5-7}
   Phase & MACs & FLOPs & Lat. & MACs & FLOPs & Lat. \\
\midrule
Training & 31.81M & 63.62M & 1.28ms & 0.358M & 0.716M & 0.16ms \\
Inference & 0 & 0 & 0 & 0 & 0 & 0 \\
\bottomrule
\end{tabular}}
\caption{\color{red}Code-derived computational complexity for deterministic chaotic objectives per 200-frame utterance. These modules are exclusively active during the training phase. Here, MACs, FLOPs are in Millions and Lat. = Latency in Milliseconds.}
\label{tab:overhead_complexity_analysis}
\end{table}

In addition, while our localized {GLARE} attention improves parameter efficiency by $73.49\%$, the architecture employs a fixed sliding window ($k=32$) and relative positional biases. This localization enables real-time inference (RTF: $0.0057$) but may restrict the model's capacity to learn global prosodic variations across extremely long phonetic sequences.} 

\section{Potential Risks/Ethical Considerations} 

While the intention of designing the CGS-ETS model is for speech restoration research purposes, it can be misused for potential secret eavesdropping, impersonation, or deepfake audio generation. This model has the potential for severe privacy and security risks. To avoid these, we have to be very careful to ensure transparency, protect consent, and always follow guidelines.

\bibliography{custom}

\appendix

\section{Appendix}
\label{sec:appendix}

\subsection{Convolutional Feature Extraction (CFE)}
\label{subsec:CFE_appendix}

CFE block comprises three residual blocks following \cite{gaddy-klein-2021-improved}. Each block comprises two 1D convolutional (Conv-1D) layers (kernel size=3) interspersed with ReLU activations, augmented by a parameterized skip connection. The first Conv-1D downsamples inputs by $2\times$ (stride=2) while the second one maintains the same resolution (stride=1). The skip connection utilizes a matching projection as the first Conv-1D (stride=2) to guarantee dimensional congruency for the residual connection. A uniform channel capacity of $C_D=512$ is rigorously maintained across all layers. \textit{This streamlined configuration is designed to capture the rapid, dynamic EMG patterns of facial articulation, which achieves superior phonetic decoding performance with a considerably reduced parameter footprint.}

\subsection{\color{red}Activation Functions}
\label{appendix:activation_function}
{\color{red}Table \ref{tab:activation_function_2} presents the complete ablation results on activation functions for different blocks of CGS-ETS (see Section \ref{subsubsec:activation_function} for detailed discussion).}

\subsection{\color{red}GLARE Attention Depth}
\label{appendix:glare_attention}
{\color{red}Table~\ref{tab:glare_attention_2} details the comprehensive GLARE attention layer depths ablation study. A full discussion of these findings is available in Section~\ref{subsubsec:glare_attention}.}

\subsection{\color{red}Different Attention Mechanisms}
\label{appendix:swa_ablation}
{\color{red}The comprehensive ablation study on attention mechanisms across various layer depths is summarized in Table~\ref{tab:Swa_ablation_2}. A detailed discussion of these results follows in Section~\ref{subsubsec:Swa_ablation}}

\begin{table*}[t!] 
\centering
\color{red}
\Large
\renewcommand{\arraystretch}{1.0} 
\resizebox{0.8\textwidth}{!}{
\begin{tabular}{@{} l l l l c c c c c c c c c @{}}
\toprule
\textbf{} & \textbf{CFE} & \textbf{Mam.} & \textbf{FFN} & 
\textbf{L$\downarrow$} & \textbf{ST$\uparrow$} & \textbf{P$\uparrow$} & 
\textbf{SD$\uparrow$} & \textbf{N$\uparrow$} & \textbf{W$\downarrow$} & 
\textbf{IT$\downarrow$} & \textbf{FLOPs$\downarrow$} & \textbf{RTF$\downarrow$} \\
\midrule
$P_0$ & ReLU & SiLU & SwiGLU & 1.10 & 0.59 & 1.16 & -31.47 & 3.25 & 51.16\% & 7.12ms & 1.038G & 0.0075 \\
$P_1$ & ReLU & SiLU & GeLU & 1.09 & 0.60 & 1.18 & -31.30 & 3.28 & 47.06\% & 5.92ms & 929.76M & 0.0059 \\
$P_2$ & LyRIC & SiLU & GeLU & 1.16 & 0.54 & 1.13 & -33.22 & 2.90 & 71.86\% & 5.92ms & 929.76M & 0.0059 \\
$P_3$ & ReLU & SiLU & LyRIC & 1.10 & 0.59 & 1.17 & -31.85 & 3.23 & 52.63\% & 5.70ms & 928.43M & 0.0057 \\
$P_4$ & ReLU & LyRIC & SwiGLU & 1.08 & 0.61 & 1.19 & -32.63 & 3.30 & 45.71\% & 6.21ms & 1.035G & 0.0062 \\
\midrule
{$P_5$} & {ReLU} & {LyRIC} & {GeLU} & 
{1.08} & {0.61} & {1.19} & {-32.61} & 
{3.33} & {41.36\%} & {5.61ms} & {926.55M} &{0.0057} \\
\bottomrule
\end{tabular}}
\caption{\color{red}Ablation Study on Activation Functions.}
\label{tab:activation_function_2}
\end{table*}

\begin{table*}[t!]
\centering
\color{red}
\Large
\renewcommand{\arraystretch}{1.0}
\resizebox{0.8\textwidth}{!}{%
\begin{tabular}{ll c c c c c c c c c c}
\toprule
\textbf{} & \textbf{Layers} & 
\textbf{L$\downarrow$} & \textbf{ST$\uparrow$} & 
\textbf{P$\uparrow$} & \textbf{SD$\uparrow$} & 
\textbf{N$\uparrow$} & \textbf{WER$\downarrow$} & 
\textbf{Para.$\downarrow$} & \textbf{IT$\downarrow$} & 
\textbf{FLOPs$\downarrow$} & \textbf{RTF$\downarrow$} \\
\midrule
$P_0$ & No Attention & 1.15 & 0.53 & 1.14 & -34.42 & 2.91 & 73.22\% & 12.68M & 4.23ms & 848.33M & 0.0046 \\
$P_1$ & $L=1,2$ & 1.08 & 0.60 & 1.18 & -32.41 & 3.30 & 46.14\% & 14.34M & 5.91ms & 926.55M & 0.0059 \\
$P_2$ & $L=3,4$ & 1.10 & 0.59 & 1.17 & -31.96 & 3.25 & 50.25\% & 14.34M & 5.97ms & 926.55M & 0.0059 \\
$P_3$ & $L=1,4$ & 1.10 & 0.59 & 1.17 & -33.54 & 3.30 & 49.01\% & 14.34M & 5.91ms & 926.55M & 0.0061 \\
$P_4$ & $L=1,2,3$ & 1.08 & 0.61 & 1.19 & -33.31 & 3.26 & 46.38\% & 15.05M & 6.86ms & 966.45M & 0.0068 \\
$P_5$ & $L=2,3,4$ & 1.09 & 0.60 & 1.19 & -32.81 & 3.28 & 44.91\% & 15.05M & 7.11ms & 966.45M & 0.0071 \\
$P_6$ & $L=1,2,3,4$ & 1.08 & 0.61 & 1.18 & -31.94 & 3.24 & 44.56\% & 15.83M & 7.58ms & 1.006G & 0.0077 \\
\midrule
\textbf{$P_7$} & {$L=1,3$} & 
{1.08} & {0.61} & {1.19} & 
{-32.61} & {3.33} & {41.36\%} & 
{14.34M} & {5.61ms} & {926.55M} & {0.0057} \\
\bottomrule
\end{tabular}}
\caption{\color{red}Ablation Study on GLARE Attention.}
\label{tab:glare_attention_2}
\end{table*}

\begin{table*}[t!] 
\centering
\color{red}
\Large
\renewcommand{\arraystretch}{1.0}
\resizebox{0.8\textwidth}{!}{
\begin{tabular}{l l c c c c c c c c c c}
\toprule
\textbf{} & \textbf{Attention Type} & \textbf{L$\downarrow$} & \textbf{ST$\uparrow$} & \textbf{P$\uparrow$} & \textbf{SD$\uparrow$} & \textbf{N$\uparrow$} & \textbf{W$\downarrow$} & \textbf{Enc.$\downarrow$} & \textbf{IT$\downarrow$} & \textbf{FLOPs$\downarrow$} & \textbf{RTF$\downarrow$} \\
\midrule
$P_0$ & MHA (All Layers) & 1.08 & 0.61 & 1.19 & -32.43 & 3.30 & 43.32\% & 14.49M & 7.42ms & 1.166G & 0.0074 \\
$P_1$ & MHA ($L=1,3$) & 1.08 & 0.61 & 1.18 & -32.36 & 3.27 & 44.73\% & 12.39M & 5.92ms & 1.061G & 0.0059 \\
$P_2$ & MQA (All Layers) & 1.07 & 0.60 & 1.18 & -31.69 & 3.33 & 44.85\% & 10.56M & 7.10ms & 966.50M & 0.0071 \\
$P_3$ & MQA ($L=1,3$) & 1.08 & 0.60 & 1.18 & -31.81 & 3.31 & 44.90\% & 9.38M & 5.69ms & 907.41M & 0.0058 \\
\midrule
\textbf{$P_4$} & {GLARE ($L=1,3$)} & {1.08} & {0.61} & {1.19} & {-32.61} & {3.33} & {41.36\%} & {9.85M} & {5.61ms} & {926.55M} & {0.0057} \\
\bottomrule
\end{tabular}}
\caption{\color{red}Ablation Study on Different Attentions}
\label{tab:Swa_ablation_2}
\end{table*}

\subsection{\color{red}Channel Dimension of CGS-ETS}
\label{subsubsec:channel_dim}

\begin{table*}[t!] 
\centering
\color{red}
\Large
\renewcommand{\arraystretch}{1.0} 
\resizebox{0.8\textwidth}{!}{%
\begin{tabular}{l l c c c c c c c c c c}
\toprule
\textbf{} & \textbf{Dimension} & 
\textbf{L$\downarrow$} & \textbf{ST$\uparrow$} & 
\textbf{P$\uparrow$} & \textbf{SD$\uparrow$} & 
\textbf{N$\uparrow$} & \textbf{W$\downarrow$} & 
\textbf{Pr$\downarrow$} & \textbf{IT$\downarrow$} & 
\textbf{FLOPs$\downarrow$} & \textbf{RTF$\downarrow$} \\
\midrule
$P_0$ & $C_D=128$  & 1.13 & 0.57 & 1.15 & -34.35 & 3.09 & 58.88\% & 0.91M & 5.53ms & 59.64M & 0.0055 \\
$P_1$ & $C_D=256$  & 1.11 & 0.59 & 1.16 & -33.09 & 3.20 & 52.14\% & 3.59M & 5.59ms & 234.030M & 0.0056 \\
$P_2$ & $C_D=1024$ & 1.08 & 0.61 & 1.19 & -32.67 & 3.34 & 45.10\% & 56.82M & 15.72ms & 3.690G & 0.0088 \\
\midrule
\textbf{$P_3$} & {$C_D=512$} & 
{1.08} & {0.61} & {1.19} & 
{-32.61} & {3.33} & {41.36\%} & 
{14.34M} & {5.61ms} & 
{926.55M} & {0.0057} \\
\bottomrule
\end{tabular}}
\caption{\color{red}Ablation Study on Channel Dimension}
\label{tab:channel_dim}
\end{table*}

{\color{red}
Table~\ref{tab:channel_dim} ablates internal channel capacity ($C_D$) for continuous EMG modeling. Small capacities ($P_0, P_1 \leq 256$) are lightweight but cause phonetic collapse (WER $>52.14\%$, NISQA $\leq3.20$). Conversely, oversizing to $C_D=1024$ ($P_2$) overfits, stalling WER at 45.10\% while drastically inflating compute (56.82M parameters; 3.69G FLOPs). The proposed $C_D=512$ ($P_3$) delivers the optimal trade-off: achieving the lowest WER (41.36\%) and peak spectral fidelity (LSD: 1.08, SD: -32.61) while slashing $P_2$'s footprint by 75\% to maintain real-time feasibility (IT: 5.61ms; RTF: 0.0057).
}

\subsection{\color{red}Layer Count of GLAME}
\label{subsec:Layer_Count}

{\color{red}
Table~\ref{tab:Layer_Count} evaluates the hierarchical depth ($L_c$) of the GLAME's constitutional blocks. A shallow $L_c=2$ topology ($P_0$) yields a lightweight module (5.67M params) but underfits coarticulatory transitions, collapsing WER to 55.94\%. Extending to $L_c=5$ ($P_2$) minimizes WER (41.91\%) but violates low-latency constraints through excessive computational bloat (14.18M encoder params; $>$1.1G FLOPs) and degrades perceptual naturalness (NISQA: 3.25), indicating acoustic over-smoothing. The optimal configuration ($P_3$, $L_c=4$) balances hierarchical capacity with compression, saturating spectral metrics (STOI: 0.61; PESQ: 1.19) and achieving the highest perceptual fidelity (NISQA: 3.33) and lowest WER (41.36\%) while maintaining sub-10M parameters (9.85M) and real-time viability (RTF: 0.0057).}

\begin{table*}[t!] 
\centering
\color{red}
\Large
\renewcommand{\arraystretch}{1.0} 
\resizebox{0.8\textwidth}{!}{%
\begin{tabular}{ll c c c c c c c c c c}
\toprule
\textbf{} & \textbf{Count} & 
\textbf{L$\downarrow$} & \textbf{ST$\uparrow$} & 
\textbf{P$\uparrow$} & \textbf{SD$\uparrow$} & 
\textbf{N$\uparrow$} & \textbf{W$\downarrow$} & 
\textbf{Enc.$\downarrow$} & 
\textbf{IT$\downarrow$} & \textbf{FLOPs$\downarrow$} & 
\textbf{RTF$\downarrow$} \\
\midrule
$P_0$ & $L_c=2$ & 1.10 & 0.58 & 1.16 & -32.75 & 3.20 & 55.94\% & 5.67M & 4.52ms & 721.48M & 0.0045 \\
$P_1$ & $L_c=3$ & 1.08 & 0.60 & 1.18 & -31.45 & 3.33 & 45.18\% & 8.51M & 5.37ms & 863.65M & 0.0057 \\
$P_2$ & $L_c=5$ & 1.08 & 0.61 & 1.19 & -33.06 & 3.25 & 41.91\% & 14.18M & 8.50ms & 1.148G & 0.0091 \\
\midrule
\textbf{$P_3$} & {$L_c=4$} & 
{1.08} & {0.61} & {1.19} & 
{-32.61} & {3.33} & {41.36\%} & 
{9.85M} & {5.61ms} & 
{926.55M} & {0.0057} \\
\bottomrule
\end{tabular}}
\caption{\color{red}Ablation Study on Layer Count of GLAME}
\label{tab:Layer_Count}
\end{table*}

{\color{red}
\subsection{Disentangling the Contributions of CGS-ETS Components}
\label{appendix:disentanglement}

To address potential concerns regarding the architectural complexity of the proposed CGS-ETS model, this section rigorously isolates and quantifies the independent contributions of our novel mechanisms: the GLAME encoder, the chaotic loss functions (LER and MSDFA), the LyRIC activation function, and the GLARE attention module. Our comprehensive ablation studies demonstrate that no single component is redundantly layered; rather, each explicitly targets a distinct representational or computational bottleneck in ETS synthesis. Table~\ref{tab:master_ablation_summary} provides a consolidated overview of these marginal utilities, demonstrating the specific performance and efficiency gains relative to standard baselines. Notably, the table captures both the macro-level efficiency of the proposed GLAME encoder and the micro-level contributions of its internal components. The detailed logic and analysis for each component are discussed below.

\begin{table*}[ht!]
    \centering
    \color{red}
    \setlength{\tabcolsep}{3.5pt} 
    \footnotesize
    \setlength{\tabcolsep}{3pt} 
    \renewcommand{\arraystretch}{1.0}
    \resizebox{1\textwidth}{!}{%
    \begin{tabular}{@{} l | c | l | l | l | l @{}}
    \toprule
    \textbf{Component} & \textbf{Ref.} & \textbf{ Bottleneck} & \textbf{Without Component} & \textbf{With Proposed Component} & \textbf{Efficiency Gain} \\
    \midrule
    
    \textbf{GLAME} & \multirow{2}{*}{Tab. \ref{tab:computational_complexity}} & Macro & \cite{gaddy-klein-2021-improved} Encoder & \textbf{GLAME} & {-34.18M Params} \\
    (Encoder) & & Architectural Bloat & (Enc: 44.03M, FLOPs: 2400M) & (Enc: 9.85M, FLOPs: 926.6M) & {-1473.4M FLOPs}\\
    \midrule
    
    \textbf{Chaos Losses} & \multirow{2}{*}{Tab. \ref{tab:loss_ablation}} & Acoustic Fidelity, & w/o LER+MSDFA ($P_1$) & \textbf{w/ LER+MSDFA} ($P_4$) & \multirow{2}{*}{N/A} \\
    (LER + MSDFA) & & Phonetic Accuracy & (WER: 51.56\%, LSD: 1.10) & (WER: 41.36\%, LSD: 1.08) & \\
    \midrule
    
    \textbf{LyRIC Activation} & \multirow{2}{*}{Tab. \ref{tab:activation_function}} & Computational & SiLU + SwiGLU ($P_0$) & \textbf{LyRIC + GeLU} ($P_5$) & {-111.4M FLOPs} \\
    (in Mamba block) & & Cost, Latency & (WER: 51.16\%, FLOPs: 1038M) & (WER: 41.36\%, FLOPs: 926.6M) & {-1.51ms Latency} \\
    \midrule
    
    \textbf{GLARE Attention} & \multirow{2}{*}{Tab. \ref{tab:Swa_ablation}} & Model Size, & MHA All Layers ($P_0$) & \textbf{GLARE $L \in \{1,3\}$} ($P_4$) & {-4.64M Params} \\
    (Alternating Layers) & & Inference Speed & (WER: 43.32\%, Enc: 14.49M) & (WER: 41.36\%, Enc: 9.85M) &{-239.4M FLOPs} \\
    
    \bottomrule
    \end{tabular}%
    }
    \caption{\normalfont \color{red} Consolidated Summary of Individual Component Contributions. This table highlights the isolated impact of each proposed mechanism compared to its standard counterpart.}
    \label{tab:master_ablation_summary}
\end{table*}

\subsubsection{The Macro-Architectural Role of the GLAME Encoder} 
While the subsequent subsections detail the internal mechanisms of our model, Table~\ref{tab:computational_complexity} highlights the overarching structural contribution of the GLAME encoder itself. Standard baseline encoders \cite{gaddy-klein-2021-improved} suffer from severe architectural bloat, requiring 44.03M parameters and 2400M FLOPs to process EMG features. By housing our state-space models within the GLAME framework, we slash the encoder size by \textbf{77.6\%} (down to 9.85M parameters) and reduce the computational overhead by \textbf{61.4\%} (down to 926.6M FLOPs). This macro-level compression is precisely what enables CGS-ETS to maintain highly efficient real-time synthesis capabilities (RTF: 0.0057) while simultaneously delivering the SOTA acoustic fidelity demonstrated in Table~\ref{tab:comparison}.

\subsubsection{The Role of Chaos-Informed Losses (LER \& MSDFA)} 

As detailed in Table~\ref{tab:loss_ablation}, the chaos-informed loss objectives act as the primary drivers of downstream intelligibility and macro-level acoustic fidelity. Relying solely on standard reconstruction losses (w/o LER+MSDFA, $P_1$) yields a suboptimal Word Error Rate (WER) of 51.56\%. When isolated, these two objectives address different temporal scales of the chaotic speech signal:
\begin{itemize}
    \item \textbf{Localized Gradient Regularization (LER):} Applying LER independently ($P_2$) stabilizes short-term, rapid signal fluctuations, dropping the WER by 3.22\% (down to 48.34\%).
    \item \textbf{Macro-Level Fractal Correlations (MSDFA):} Applying MSDFA independently ($P_3$) captures long-range dependencies across phonemes, dropping the WER by 5.91\% (down to 45.65\%) and boosting perceptual naturalness.
\end{itemize}
Their synergistic application ($P_4$) achieves the optimal theoretical bound, constraining the deterministic chaotic invariants of human phonation to minimize WER to 41.36\% while maximizing absolute acoustic fidelity (LSD: 1.08, STOI: 0.61).

\subsubsection{The Role of the LyRIC Activation Function}

While the chaos losses enforce physical constraints, the network requires sufficient inductive bias to map these constraints efficiently. Table~\ref{tab:activation_function} illustrates that standard activation routing (SiLU and SwiGLU, $P_0$) struggles with this mapping, incurring a severe computational penalty of 1038M FLOPs alongside a high WER of 51.16\%. By anchoring our Lyapunov-inspired LyRIC activation exclusively within the Mamba state-space block, and utilizing GeLU in the terminal FFN ($P_5$), we resolve this computational-acoustic trade-off. This specific configuration successfully internalizes the chaotic dynamics, reducing the computational footprint by over 111M FLOPs (down to 926.6M FLOPs) while simultaneously driving phonetic accuracy to its peak (41.36\% WER). Placing LyRIC elsewhere (e.g., in the CFE or FFN) actively destabilizes the representation, proving its highly specific utility within the state-space framework.

\subsubsection{The Role of GLARE Attention} 

Finally, Table~\ref{tab:Swa_ablation} isolates the structural efficiency provided by the GLARE module. Standard Multi-Head Attention (MHA) across all layers ($P_0$) suffers from extreme representational redundancy, requiring 14.49M encoder parameters, 1166M FLOPs, and 7.42ms of latency to achieve a 43.32\% WER. Simply truncating MHA or switching to MQA variants ($P_1, P_2, P_3$) successfully reduces the parameter count but severely degrades intelligibility (pushing WER back up to $\sim$45\%).

By replacing MHA with our localized GLARE attention at alternating layers ($L \in \{1, 3\}$, $P_4$), we solve this redundancy. GLARE reduces encoder parameters by 32\% (down to 9.85M) and inference latency by 24\% (down to 5.61ms). This massive compression acts as a powerful regularizer, filtering out acoustic noise and actually \textit{improving} downstream intelligibility (dropping WER to 41.36\%).

The integration of these moving parts is a highly deliberate pipeline. The chaotic losses (LER and MSDFA) define the physical targets of human phonation, while the GLAME architecture, LyRIC activation, and GLARE attention provide the specific, parameter-efficient inductive biases required to achieve those targets in real-time.
}

\subsection{EMG as A Chaotic Dynamical System}
\label{subsec:Chaotic_PropertiesofEMG}

Articulatory EMG captures the coordinated, spatio-temporal activation of facial and laryngeal muscles during speech production \cite{gaddy-klein-2020-digital, gaddy-klein-2021-improved}. This myoelectric generation is a \emph{non-linear dynamical process characterized by deterministic chaos} \cite{padmanabhan2004nonlinear, ravier2005fractal}. The physiological ``source'' of the signal is the central neural drive, which recruits populations of motor neurons. These neurons subsequently generate asynchronous Motor Unit Action Potentials (MUAPs) across varying muscle fibers. However, before reaching the surface electrodes, these localized action potentials undergo severe non-linear spatial filtering through the visco-elastic subcutaneous tissue, skin layers, and shifting muscle geometries \cite{merletti2016surface}. The resulting surface EMG is not a simple linear summation, but rather a highly complex interference pattern characterized by amplitude modulation, quasi-periodic bursting, and phase cancellations \cite{farina2014}.

Essentially, the articulatory neuromuscular system operates through a continuous, closed-loop sensorimotor framework. Motor unit recruitment is dynamically modulated in real-time by proprioceptive feedback from muscle spindles and Golgi tendon organs \cite{todorov2004optimality}. Because the central nervous system rapidly adjusts firing rates based on this non-linear feedback, the resulting muscle contractions exhibit extreme sensitivity to initial conditions. Consequently, the EMG signal frequently undergoes bifurcations-- such as sudden motor unit synchronization during rapid coarticulation or fatigue-induced tremors—intermittently driving the signal into \emph{chaotic regimes} rigorously indicated by positive Lyapunov exponents \cite{padmanabhan2004nonlinear}. 

The bio-signal naturally alternates between quasi-stationary periods (e.g., sustained vowel postures) and highly stochastic, irregular micro-fluctuations (e.g., rapid consonant transitions) during continuous phonetic articulation. Moreover, the transient biochemical kinetics of ion channels (e.g., $Na^+$/$K^+$ gradients) and varying action potential conduction velocities introduce continuous micro-temporal jitter into the firing patterns \cite{merletti2016surface}. This creates a non-stationary, fractal-like time-series where minor physiological perturbations map to massive variations in the recorded interference pattern.

A purely linear regression model or standard sequence-to-sequence transformer basically overlooks these critical bio-physical realities \cite{gaddy-klein-2021-improved}. Conventional models mathematically average out the sub-harmonic firing frequencies, bifurcations, and aperiodic MUAP bursts by optimizing strictly for mean-squared reconstruction. In the context of ETS synthesis, this directly causes the network to output unnaturally over-smoothed, muffled acoustic representations devoid of natural prosody and crisp phonetic transitions. Therefore, to faithfully invert this bio-signal into highly intelligible speech, the predictive architecture must explicitly incorporate non-linear, chaos-inspired topological constraints (e.g., LER, MSDFA) capable of mapping both the structured macroscopic muscle gestures and the deterministic chaotic micro-structures of the electrophysiological source.

\subsection{Theoretical and Empirical Analysis of the LyRIC Activation Function}
\label{subsec:Lyric_appendix}

Lyapunov theory provides the mathematical foundation for analyzing the stability and dynamic behavior of complex nonlinear systems without requiring explicit analytical solutions to their underlying differential equations. Modeling continuous biological signals such as EMG and chaotic dynamical systems such as speech through Lyapunov theory relies on two fundamental pillars: \textbf{Equilibrium Stability} and  \textbf{Deterministic Chaos}. Equilibrium Stability theory help us design the \textbf{LyRIC activation} function, while Deterministic Chaos behavior underpins the formulation of the \textbf{LER loss}. A classical Lyapunov function in nonlinear dynamical system must satisfy the Equilibrium Stability conditions.

\textbf{Equilibrium Stability:} This method assesses whether a perturbed system will return to a  state of equilibrium. It introduces a scalar, continuously differentiable ``energy-like'' function, $V(x)$, to evaluate the system state $x$. For an equilibrium point at the origin ($x = 0$), the system is considered \textbf{asymptotically stable} if a $V(x)$ exists such 
that:
\begin{itemize}
    \item \textbf{Positive Definite:} $V(x) > 0$ for all $x \neq 0$, 
    and $V(0) = 0$. (The system has positive ``energy'' when 
    displaced).

    \item \textbf{Negative Definite Derivative:} The temporal 
    derivative along the system's trajectory, 
    $\dot{V}(x) = \nabla V(x) \cdot \dot{x}$, is strictly less than 
    zero ($\dot{V}(x) < 0$ for $x \neq 0$).

    \item \textbf{Radically Unbounded:} $V(x)\to\infty$ as $\vert\vert x\vert\vert\to\infty$ for global stability.
\end{itemize}

\subsubsection{LyRIC Activation Properties}
\label{subsubsec:lyric_properties}
In section \ref{subsec:lyric}, we derived the following equation (Equation \ref{eqn2}): $\small g(x) = x \left[ \frac{\log(\vert x \vert^\alpha + 1)}{\vert x \vert^\alpha + 1} \right]$.
From this equation, we derived our main activation function: $\small f(x) = x \vert x \vert^\alpha$.
\textbf{Log-divergence} equation (Equation~\ref{eqn2}) and \textbf{LyRIC} activation function (Equation~\ref{eqn6}) are mathematically guaranteed to preserve the following critical topological and optimization 
properties, thereby establishing Lyapunov's Equilibrium Stability theory:
 \begin{itemize}
    \item \textbf{Odd Symmetry and Origin Linearity:} The explicit multiplication by the raw input $x$ rigorously preserves the directional phase (sign) of the latent features, while smoothly degenerating to a trivial zero map ($g(0)=0$) at the origin.
    \item \textbf{Asymptotic Attenuation:} The stability normalization divisor, $\vert x \vert^\alpha + 1$, guarantees that the activation smoothly decays toward zero for excessively large inputs ($\lim_{|x| \to \infty} g(x) = 0$), acting as a powerful dynamic regularizer against outlier activations.
    \item \textbf{Adaptive Phase-Space Expansion:} The learnable parameter $\alpha$ empowers the network to dynamically optimize the rate of non-linear expansion directly from the morphological features of the physiological data such as EMG.
    \item \textbf{Deterministic Chaotic Mapping:} The core term, $\log(\vert x \vert^\alpha + 1)$, directly models the exponential divergence characteristic of chaotic dynamical systems, compelling the network to maintain acute sensitivity to critical micro-variations in the bio-signal.
    \item \textbf{Smoothness and Gradient Stability:} The function is continuously differentiable ($C^1$ smooth) across its entire domain, ensuring well-behaved, stable gradient propagation throughout deep architectural hierarchies.
 \end{itemize}

\subsubsection{Comprehensive Analysis of Log-divergence Equation}
\label{subsubsec:log-divergence}

\textbf{Visual Analysis:} Figure \ref{fig:log-div-plot} provides a visual analysis of the Lyapunov-derived Log-divergence equation (Equation \ref{eqn2}) manifold. The left panel contrasts its topology against conventional activations (ReLU, GELU, SiLU). While standard functions truncate the negative domain and exhibit unbounded linear growth for positive inputs, the Lyapunov formulation guarantees strict odd symmetry (bilateral phase preservation) and imposes sub-linear asymptotic attenuation, bounding extreme activations to prevent gradient explosion. The right panel empirically demonstrates the parametric influence of the expansion coefficient, $\alpha$. Variations in $\alpha$ dynamically modulate the non-linear expansion rate with higher values (e.g., $\alpha=2.0$) enforcing tighter asymptotic bounds while mathematically preserving zero-centered topological stability ($f(0)=0$) across all configurations.

\begin{figure}[ht!]
    \centering
    \includegraphics[width=0.4\textwidth]{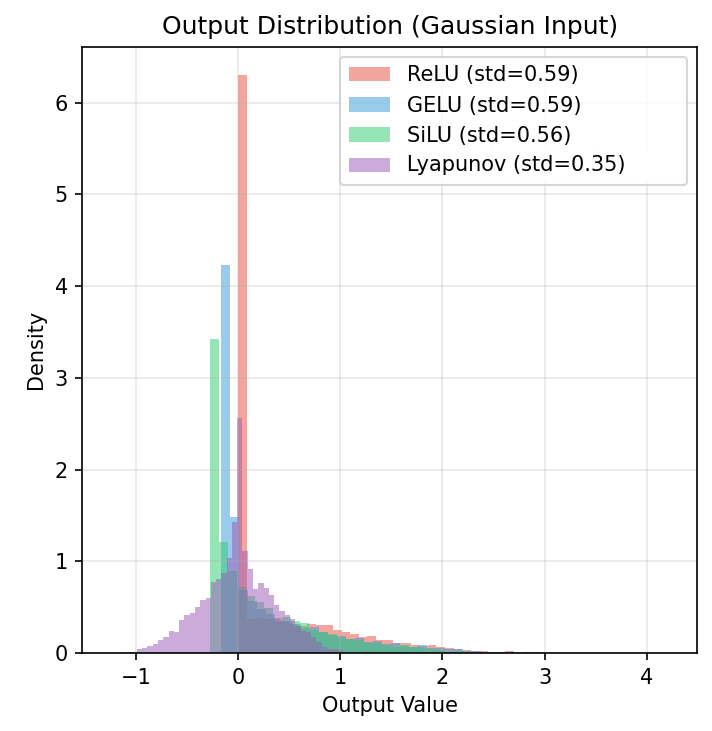}
    \caption{Output Density Distributions of ReLU, GELU, SiLU against Log-divergence Equation (Equation \ref{eqn6})}
    \label{fig:distribution_eqn6}
\end{figure}

\textbf{Output Density Distributions:} Figure \ref{fig:distribution_eqn6} contrasts the output density distributions of standard activation functions (ReLU, GELU, SiLU) against Log-divergence equation (Equation \ref{eqn6}) under a standard normal input sequence ($x \sim \mathcal{N}(0,1)$). As illustrated, conventional activations inherently suppress negative inputs, culminating in a severe density spike at the origin. This zero-state saturation not only induces catastrophic representational collapse (the ``dying neuron'' problem) but also artificially inflates the output variance ($\sigma \in[0.56, 0.59]$) through extreme distributional imbalance. The Lyapunov derived function (Equation  \ref{eqn6}) preserves bilateral phase geometry. By seamlessly extending into the negative domain, it strictly prevents zero-collapse, ensuring continuous gradient flow. Its considerably lower variance ($\sigma = 0.35$) is a direct mathematical manifestation of Lyapunov \textit{Equilibrium Stability}. This function rigorously bounds the state-space trajectory by guaranteeing that latent representations neither collapse to absolute zero ($V(x) > 0$ for $x \neq 0$) nor exponentially explode ($\dot{V}(x) < 0$). \textit{This theoretical constraint yields a tightly controlled, zero-centered distributional footprint, ensuring the uniform and mathematically stable signal utilization required to model the deterministic chaos of continuous EMG sequences.}

\textbf{Computational Profiling:} Table~\ref{tab:flops_activation} quantifies the element-wise computational overhead of the exact log-divergence 
formulation (Equation~\ref{eqn6}) against standard nonlinear activations. Hardware-optimized functions such as ReLU ($P_0$), GELU ($P_1$), and SiLU ($P_2$) require minimal floating-point  operations (1--22 FLOPs/element). However, direct evaluation of 
the nested logarithmic and fractional terms in the exact  formulation ($P_3$) incurs 47 FLOPs per element. This severe latency bottleneck-demanding more than double the computational burden of SiLU motivates and mathematically justifies our prior derivation of the streamlined LyRIC polynomial approximation (Equation \ref{eqn6}) which ensures robust chaotic dynamic modeling without compromising the high-throughput inference requirements of real-time ETS synthesis.

\begin{table}[ht!]
\centering
\resizebox{0.6\columnwidth}{!}{%
\begin{tabular}{l l c}
\toprule
\textbf{} & \textbf{Function} & \textbf{FLOPs/element$\downarrow$} \\
\midrule
$P_0$ & ReLU & 1 \\
$P_1$ & GeLU & 14 \\
$P_2$ & SiLU & 22 \\
$P_3$ & Log-divergence & 47 \\
\bottomrule
\end{tabular}}
\caption{Comparison of Computational Profiling among ReLU, GeLU, SiLU, and Log-divergence Equation (see Equation~\ref{eqn6})}

\label{tab:flops_activation}
\end{table}

\subsubsection{Comprehensive Analysis of LyRIC Activation}
\label{subsubsec:lyric_activation_detail}
\textbf{Motivation: }As quantified in Table \ref{tab:flops_activation}, the direct evaluation of the nested logarithmic and fractional operators in Equation \ref{eqn6} imposes a prohibitive FLOPs penalty relative to hardware-optimized baselines (ReLU, GELU, SiLU). To resolve this computational bottleneck, we derive Equation \ref{eqn6} which is a highly efficient, lower-bound polynomial approximation formulated through truncated Maclaurin and geometric series expansions. This mathematically rigorous simplification neutralizes the algorithmic overhead while faithfully preserving the deterministic chaotic expansion properties and topological stability of the original continuous manifold. As mentioned in Table~\ref{tab:flops_lyric}, we set $\alpha{=}1$ for LyRIC to minimize computational cost (1 FLOP/element). 

\begin{figure}[ht!]
    \centering
    \includegraphics[width=0.4\textwidth]{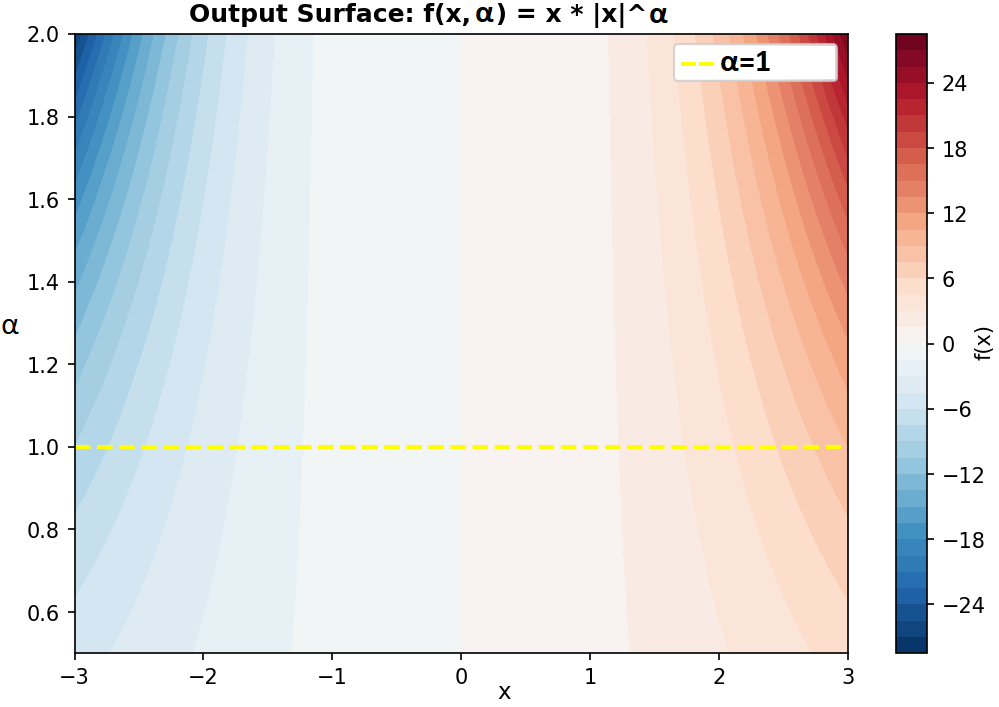}
    \vspace{-1em}
    \caption{Topological Sensitivity Analysis of the LyRIC Activation Surface ($f(x, \alpha) = x|x|^{\alpha}$).}
    \label{fig:lyric_phase}
\end{figure}

\textbf{Topological Sensitivity Analysis:} The heat map in Figure \ref{fig:lyric_phase} visualizes the output magnitude across the input ($x$) and exponent ($\alpha$) manifold. The surface exhibits odd symmetry (bilateral phase preservation) and high topological continuity across the entire input domain. The dashed yellow iso-line ($\alpha = 1.0$) identifies the region of optimal gradient stability, where the function achieves a balance between deterministic nonlinear expansion and controlled asymptotic attenuation. The absence of sharp discontinuities or saturation plateaus confirms the activation's suitability for preserving high-entropy, chaotic EMG signal microstructures. To evaluate the operational stability of the proposed LyRIC 
activation function, we perform a sensitivity analysis on its output surface relative to the expansion coefficient, $\alpha \in [0.6, 2.0]$. As illustrated in Figure~\ref{fig:lyric_phase}, the activation function maintains a continuous, smooth gradient landscape across the entire input-exponent manifold. Unlike standard saturating activations that induce hard nonlinearities, the LyRIC surface exhibits gradual, predictable 
transitions. This is critical for ETS synthesis, as it ensures that the model remains sensitive to the non-stationary fluctuations of articulatory EMG signals rather than suppressing them via premature saturation. The observed heat map color-coding --- transitioning uniformly from dark blue (negative magnitude) to white (origin) to dark red (positive magnitude) --- demonstrates the function's strict odd symmetry, ensuring that bidirectional neuromuscular activations are mapped with uniform topological fidelity. The iso-line at $\alpha = 1.0$ (highlighted) represents the selected operational point. This configuration acts as a stable equilibrium for the neural network, providing sufficient nonlinear expansion to model chaotic attractors while avoiding the representational instability associated with higher-order exponentiation ($> 1.5$). Therefore, the LyRIC activation provides the representational bandwidth to decode the rapid, aperiodic micro-fluctuations characteristic of deterministic chaotic dynamics in human motor control, while maintaining a minimal computational footprint of 1~FLOP/element.

\begin{figure}[ht!]
    \centering
    \includegraphics[width=0.45\textwidth]{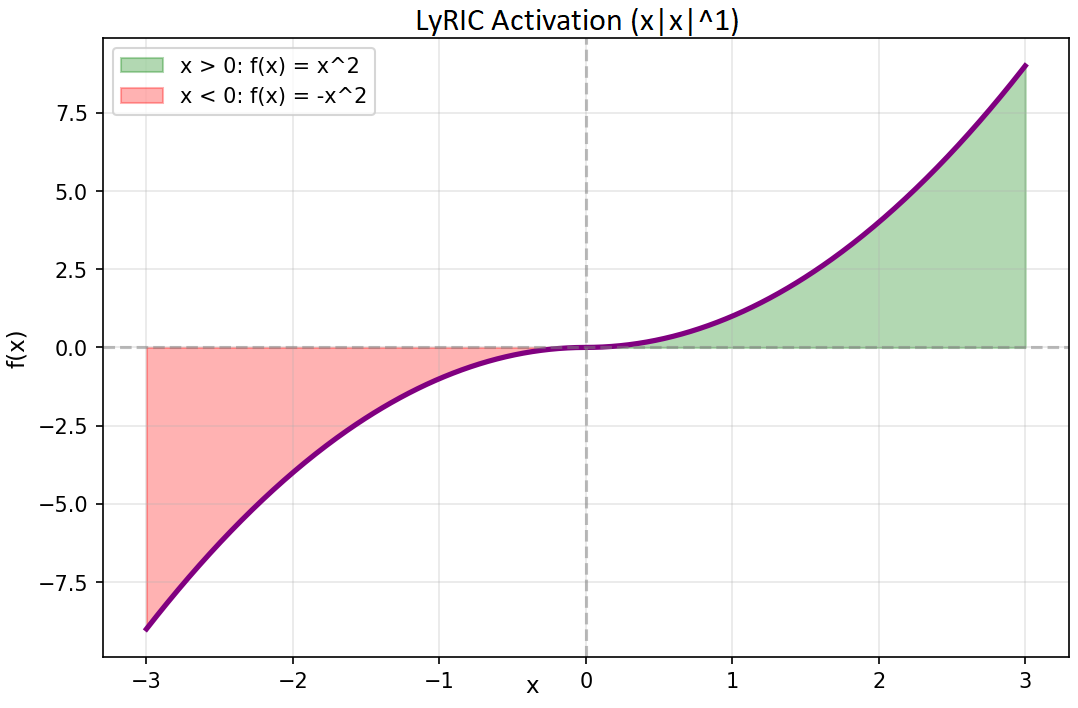}
    \vspace{-1em}
    \caption{Geometric Representation of the LyRIC Activation Function at the Optimal Expansion Coefficient $\alpha=1.0$.}
    \label{fig:lyric_geometry}
\end{figure}

\textbf{Geometric Representation:} 
The LyRIC Activation function can be reduced to a signed square topology ($f(x) = \text{sign}(x) \cdot x^2$). Figure \ref{fig:lyric_geometry} demonstrates continuous differentiability and bilateral phase preservation. As derived in Section \ref{subsec:lyric} (Equation \ref{eqn6}), our LyRIC formulation simplifies to a signed square topology when $\alpha=1.0$, expressed as $f(x) = \text{sign}(x) \cdot x^2$. This topology provides a fundamental advantage over standard rectifying activations: it eliminates the zero-state saturation zone ($x < 0$). LyRIC maintains the full representational bandwidth of the EMG source signal by mapping negative signal fluctuations to negative outputs. This bilateral sensitivity ensures that the encoder remains explicitly responsive to the rapidly shifting phase transitions and aperiodic bursts that characterize the chaotic dynamics of neuromuscular motor unit recruitment.

\begin{figure}[ht!]
    \centering
    \includegraphics[width=0.45\textwidth]{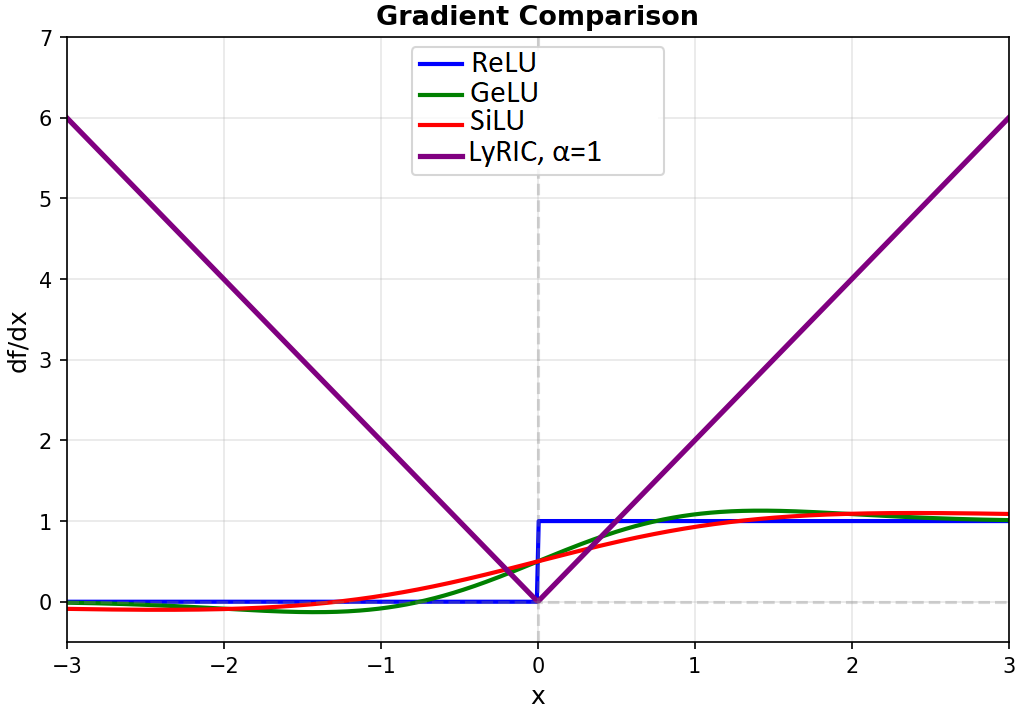}
    \vspace{-1em}
    \caption{Gradient Landscape Analysis of ReLU, GeLU, SiLU, and LyRIC ($\alpha=1$).}
    \label{fig:lyric_gradient}
\end{figure}

\begin{figure}[ht!]
    \centering
    \includegraphics[width=0.45\textwidth]{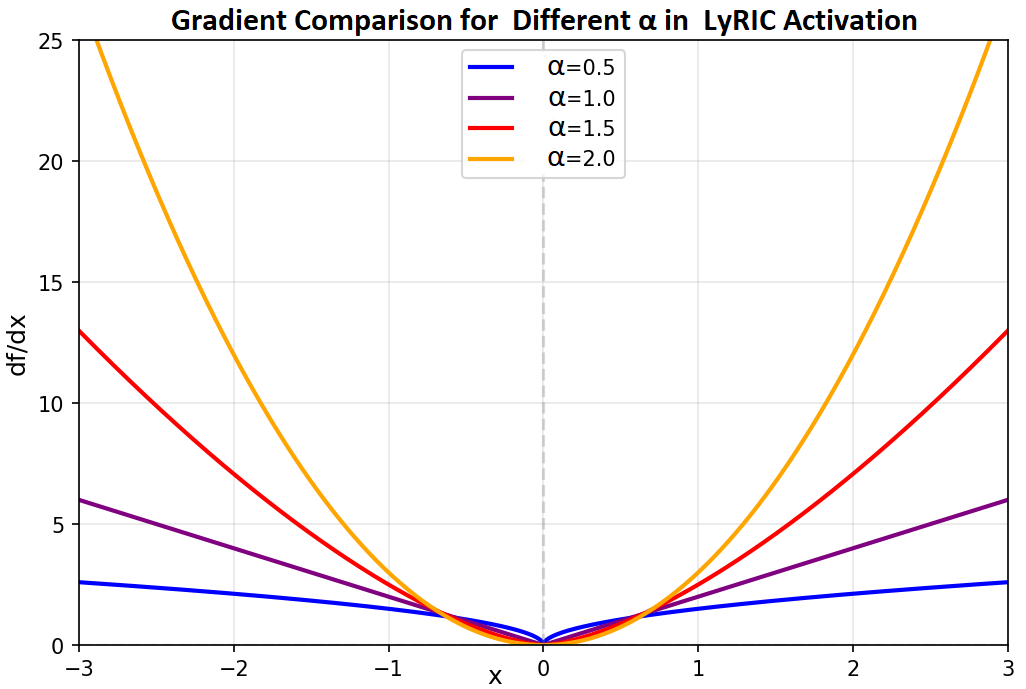}
    \vspace{-1em}
    \caption{Parametric Influence of the Expansion Coefficient ($\alpha$) on the Gradient Landscape of the LyRIC Activation.}
    \label{fig:lyric_diffgradient}
\end{figure}

\textbf{Gradient Landscape Analysis:} The \textit{LyRIC} activation $f(x) = x|x|$ has gradient $f'(x) = 2|x|$, which scales proportionally with input magnitude, as shown in Figure~\ref{fig:lyric_gradient}. Unlike saturating activations (ReLU, GELU, SiLU) whose gradients converge to 1.0 or 0.0, LyRIC exhibits adaptive gradient expansion, enabling the network to propagate high-magnitude signals without saturation artifacts. This nonsaturating property is mathematically analogous to a positive Lyapunov exponent, ensuring the model retains sensitivity to the high-variance, aperiodic bursts characteristic of chaotic EMG articulatory dynamics throughout the GLAME. Figure~\ref{fig:lyric_diffgradient} illustrates how $\alpha$ controls gradient gain across the latent manifold to analyze the influence of expansion rate $\alpha$ on the gradient function $f'(x) = (1+\alpha)|x|^\alpha$. Lower values ($\alpha = 0.5$) produce a flat gradient profile, risking representational underfitting for subtle EMG fluctuations, while higher values ($\alpha > 1.0$) accelerate gradient magnitude which potentially induce optimization instability. The configuration $\alpha{=}1.0$ provides a balanced, linear expansion rate that preserves the high-entropy microstructure of the EMG signal while maintaining stable gradient flow throughout the GLAME.

\subsection{Theoretical Analysis of the LER Loss}
\label{subsec:LER_appendix}
Deterministic Chaos theory \cite{strogatz1994, lorenz1963} of Lyapunov in nonlinear dynamical systems motivates the formulation of the \textbf{LER loss}. It describes complex, non-periodic behavior in nonlinear dynamical systems that are highly sensitive to initial conditions. While such systems are governed by deterministic laws, their long-term evolution is inherently unpredictable due to the exponential amplification of infinitesimal perturbations.

\textbf{Deterministic Chaos:} In Deterministic Chaos theory, the fundamental metric for quantifying the chaotic behavior is the \textit{Lyapunov Exponent} ($\lambda$), which measures the mean exponential rate $|\delta x(t)|$ of trajectory divergence in state-space. The spectrum of $\lambda$ classifies the system dynamics: $\lambda < 0$ denotes a stable, over-damped system converging to a fixed point; $\lambda = 0$ corresponds to marginally stable limit cycles or periodic oscillators; and $\lambda > 0$ serves as the definitive indicator of deterministic chaos. In such chaotic regimes, microscopic variations in the initial state map to macroscopic, divergent variations in the system's output-- \textit{a hallmark of human phonation and neuromuscular motor control}.

\textbf{Deterministic Chaos for Designing LER Loss:} Conventional ETS systems optimize for frame-level reconstruction losses (e.g., L1 or MSE loss), which mathematically enforces a linear, stable convergence ($\lambda \le 0$). This objective forces the network to eliminate natural jitter, aperiodic bursts, and rapid state transitions, resulting in unnaturally smoothed acoustic prosody. To counter this, we introduce the \textit{Lyapunov Exponent Regularization (LER) loss}, designed to preserve the chaotic manifold of the source signals. As formalized in Algorithm \ref{alg:LER}, LER operates by reconstructing the latent phase-space trajectories of both the predicted EMG synthesized features and the ground-truth targets. We utilize the Rosenstein-based divergence estimation \cite{rosenstein1993}, which computes the average logarithmic rate of separation between neighboring trajectories. By enforcing a penalty on the distance between the predicted ($\lambda_{\hat{Y}}$) and ground-truth ($\lambda_{Y}$) Lyapunov exponents, we constrain the encoder to replicate the \textit{divergence characteristics} of the source neuromuscular signals. In the context of our EMG systems, this physically grounded supervision ensures the model does not collapse into a stable equilibrium. Instead, it compels the network to represent the deterministic chaos inherent in the electrophysiological source-- particularly the sub-harmonic firing patterns, bifurcations, and timing irregularities-- which are essential for synthesizing human-like, temporally coherent speech. Therefore, the LER objective serves as a topological constraint, ensuring that the synthesized acoustic features faithfully reflect the complex, non-linear micro-structure of the underlying EMG articulatory gestures.

\subsection{Dataset Preprocessing Pipeline}
\label{sec:dataset_preprocessing} 

\subsubsection{English Corpus}
{\color{red}We utilize the established SOTA 2020 ETS English dataset \cite{gaddy-klein-2020-digital} to benchmark our proposed CGS-ETS architecture. The dataset comprises nearly 20 hours of facial EMG and parallel audio recordings from a single English speaker, bifurcated into closed-vocabulary and open-vocabulary conditions. To facilitate acoustic mapping, the corpus utilizes parallel tuples of silently mouthed EMG, vocalized EMG, and simultaneous vocalized audio. During the training phase of our model, DTW on the vocalized pairs generates alignments used to transfer audio targets to the silent EMG signals. The closed-vocabulary subset contains $\approx$56 minutes of data restricted to 67 unique words and was recorded in a single, continuous session to eliminate spatial variance in electrode placement. Conversely, the open-vocabulary condition encompasses the vast majority of the corpus, featuring open-domain sentences from public domain books recorded across multiple sessions, which deliberately introduces a realistic variance in electrode positioning. For standard evaluation, this open-vocabulary data is partitioned into an 8:1:1 split for training, validation, and testing, with strict precautions taken to ensure that the specific texts utilized in the test set are excluded from any external language model training corpora to prevent label leakage.} Prior to experimentation, we rigorously verified that the corpus is devoid of Personally Identifiable Information (PII) and any abusive or harmful content, adhering to ethical standards for bio-signal research. The corpus is completely partitioned into training, validation, and test splits following the standardized JSON indices provided in the benchmark.

\textbf{\color{red}Preprocessing:} Each utterance is contextually loaded by concatenating it with its temporally adjacent neighbors ($i-1, i, i+1$) to precisely mitigate transient boundary artifacts during digital filtering which isolates the target sequence only after continuous filtration. We suppress ambient electromagnetic interference by applying a rigorous comb notch filter at 60 Hz and its subsequent seven harmonics (up to 420 Hz). A 3rd-order Butterworth high-pass filter (2 Hz cutoff) is subsequently applied to eliminate bio-electrical baseline drift. The continuous signals are then subsampled to synchronize with the acoustic frame rate ($\approx$689 Hz). Following filtration, we extract a dense, multi-domain feature representation utilizing a sliding window (16-sample length, 6-sample hop). This results in localized time-domain statistics (mean amplitudes, RMS envelopes, zero-crossing rates of the high-frequency residual) alongside frequency-domain STFT magnitudes. The concatenated EMG features are standardized through global mean-variance normalization and tightly bounded using a scaled hyperbolic tangent projection ($8\tanh(x/8)$) to guarantee strict topological stability during gradient propagation.

\textbf{Acoustic and Linguistic Processing:} Ground-truth acoustic waveforms are dynamically loaded, collapsed from stereo to mono configurations to isolate the primary channel, and rigorously resampled to 22.05 kHz. We extract 80-dimensional mel-spectrograms to formulate the primary regression targets utilizing a 1024-point STFT, a 1024-sample Hanning window, and a 256-sample hop size, bounding the spectral representation between 0 and 8,000 Hz. Concurrently, to facilitate our auxiliary linguistic optimization, we parse forced-alignment \texttt{TextGrid} annotations to extract frame-level phonetic labels across a 48-token inventory (including silence). Finally, raw text transcripts are stripped of punctuation, lowercased, and normalized through the \texttt{JiWER} pipeline to ensure pristine linguistic mapping during multi-task supervision.

\subsubsection{\color{red} Mandarin Corpus}
{\color{red} Adapting the CGS-ETS architecture to a completely new multi-modal, 6-channel Mandarin dataset called AVE (Audio, Visual, and Electromyographic Signals) Speech \cite{11079657} required profound modifications across the entire data processing, alignment, and evaluation pipeline. It is designed to act as a comprehensive benchmark for multi-modal speech recognition. The AVE Speech corpus consists of over 55 hours of synchronized multi-modal data per modality, collected from 100 native Mandarin speakers (71 males, 29 females; aged 18--40). The dataset is built upon 100 distinct Mandarin Chinese sentences designed to cover daily communication needs. Each subject read the entire 100-sentence corpus 10 times, yielding a total of $100 \times 100 \times 10 = 100,000$ discrete utterances. Unlike the continuous 10-minute sessions utilized in the Gaddy dataset, AVE-Speech employs discrete, fixed-length trials averaging approximately 2 seconds per sentence. The synchronized modalities captured during these trials include 16 kHz audio and 6-channel EMG signals sampled natively at 1000 Hz. The dataset was downloaded using the \texttt{huggingface\_hub} Python API via the \texttt{snapshot\_download} function. The downloaded dataset was heavily nested and compressed, with individual subjects' data stored in separate \texttt{.zip} files. A custom recursive unzipping script was deployed using Python's \texttt{zipfile} and \texttt{glob} libraries to extract the \texttt{.wav} and \texttt{.mat} files in place while preserving the structural integrity of the \texttt{Train} and \texttt{Test} splits.

\textbf{Preprocessing:} Migrating from the Gaddy dataset to AVE-Speech required rectifying multiple hardcoded assumptions in the original codebase. AVE-Speech utilizes 6-channel EMG. While the dynamic feature extraction loop natively accommodated the channel reduction, the neural network's input layers required explicit reconfiguration. The CFE block's input channel size was adjusted from 8 to 6, and the linear projection layer (\texttt{emg\_prenet}) was updated to 84 ($6 \times 14$). The baseline \texttt{load\_utterance} function was engineered to load contiguous sequences by fetching \texttt{index-1} and \texttt{index+1} files to resolve boundary artifacts. Because AVE-Speech is comprised of discrete 2-second sentences, this continuous context logic caused out-of-bounds errors and was entirely excised. A critical error was initially encountered during tensor collation (\texttt{AssertionError} in \texttt{decollate\_tensor}). Initially, the native 1000 Hz AVE-Speech EMG signal was fed directly into the pipeline under the assumption that the baseline's subsampling rates were merely hardware-specific clock-drift corrections for Gaddy's equipment. However, these values are strict mathematical rate-matching conversions required by the architecture. The audio Mel-spectrograms are generated at exactly 86.13 frames per second. Since the CFE block downsamples the raw EMG sequence by a factor of 8, the raw EMG fed into the CFE block must be exactly $86.13 \times 8 \approx 689.06$ Hz. To resolve the tensor mismatch, explicit \texttt{np.interp} subsampling functions were reintroduced to perfectly align the 1000 Hz AVE-Speech signal to the 86.13 Hz STFT frame rate. Following the data pipeline restructuring, the \texttt{FeatureNormalizer} was executed from scratch. Because the scale and baseline voltages of the AVE-Speech EMG hardware differ drastically from the original corpus, calculating the exact mean and standard deviation over the new dataset was imperative to prevent gradient explosion and to save a dataset-specific \texttt{normalizers.pkl} file.

\textbf{Phoneme Labeling via Montreal Forced Aligner (MFA):} The CGS-ETS model relies on DTW combined with a phoneme-level cross-entropy loss to enforce structural alignment. However, AVE-Speech does not provide native phoneme alignments (e.g., \texttt{.TextGrid} files). We constructed a custom alignment pipeline using the Montreal Forced Aligner (MFA). The provided \texttt{phonetic\_transcription.xlsx} lacked a header row and formatted IDs irregularly (e.g., \texttt{\#66}). A custom \texttt{pandas} script was developed using Regular Expressions (\texttt{re.findall}) to extract the numeric IDs, pad them to 4-digits (e.g., \texttt{0066}), and map them to their corresponding Mandarin text inside a \texttt{transcripts.json} dictionary. A preparation script recursively traversed the \texttt{Audio} directories to generate a \texttt{.lab} text file alongside every \texttt{.wav} file containing its respective transcript. Because MFA relies on Kaldi (which poses severe dependency conflicts with PyTorch), a separate isolated Conda environment (\texttt{mfa\_env}) was created. The \texttt{mandarin\_mfa} acoustic model and pronunciation dictionary were downloaded. The aligner successfully converted the Chinese characters to Pinyin, established millisecond-level boundaries, and output \texttt{.TextGrid} files mimicking the exact nested structure of the dataset. The \texttt{TextTransform} class in the PyTorch code was subsequently overhauled to bypass English alphabet restrictions and dynamically map Chinese characters and Pinyin phonemes to integer vocabularies.

\textbf{Evaluation Metric Paradigm Shift (WER vs. CER):}
The most fundamental evaluation shift involved abandoning Word Error Rate (WER) in favor of Character Error Rate (CER). Our codebase utilizes an English ASR evaluator (\texttt{asr\_evaluation.py}) reliant on an English Wav2Vec2/DeepSpeech model. Feeding generated Mandarin audio into an English ASR resulted in phonetic hallucinations. Furthermore, standard WER relies on spaces to delimit words, which is incompatible with Mandarin text morphology. A custom evaluation pipeline was developed utilizing OpenAI's \texttt{whisper-small} model via the Hugging Face \texttt{transformers} pipeline, forced into Mandarin transcription mode. During audio processing, the evaluation script iterates through the output directory to transcribe the generated \texttt{.wav} files. To ensure data integrity, a critical bug was resolved wherein mismatched 4-digit ID padding caused the script to fetch empty ground-truth strings, which previously crashed the \texttt{jiwer} library. The script now filters empty strings, cleans all punctuation via regular expressions, and accurately maps predictions to the corresponding ground-truth arrays. Finally, to calculate CER using standard libraries, every Chinese character in both the ground-truth and predicted strings was separated by spaces before scoring.
}

\subsection{Dataset Class and DataLoader}
\label{sec:dataset_dataloader}
The aforementioned signal processing and feature extraction pipelines are formalized within a custom PyTorch \texttt{Dataset} class (\texttt{EMGDataset}). The global utterance indices are deterministically shuffled utilizing a fixed random seed upon initialization to guarantee absolute experimental reproducibility. The total dataset capacity (\texttt{\_\_len\_\_}) reflects the number of valid sequences mapped within the predefined JSON splits. During runtime, the \texttt{\_\_getitem\_\_} method dynamically retrieves the multi-modal representations which heavily leverages Least Recently Used (LRU) caching to minimize disk I/O latency. Within this method, the raw and structured EMG features are normalized and topologically bounded through scaled hyperbolic tangent projections. For silent articulation samples, the method automatically retrieves and aligns the corresponding parallel vocalized acoustic targets. To maximize PCIe data transfer rates to the GPU, all output tensors are explicitly allocated to page-locked memory (\texttt{pin\_memory()}). During the optimization phase, we instantiate a \texttt{DataLoader} governed by a custom \texttt{SizeAwareSampler}. This sophisticated sampler dynamically aggregates sequences based on a cumulative temporal frame limit (\texttt{max\_len = 256,000}) rather than utilizing naive, fixed instance-count batching. This prevents GPU Out-Of-Memory (OOM) faults across highly variable continuous speech sequences while maximizing parallel processing throughput. Finally, a custom collation function (\texttt{collate\_raw}) forcibly pads these variable-length multi-modal tensors and seamlessly routes parallel acoustic targets when processing silent EMG inputs.

\subsection{Details on Computational Complexity}
\label{append:analysisoncomputationcomplexity}

Table \ref{tab:computational_complexity} contrasts the inference-time computational complexity of our proposed CGS-ETS architecture against the SOTA baseline ($P_0$) \cite{gaddy-klein-2021-improved} on English dataset \cite{gaddy-klein-2020-digital}.

\textbf{Model Size (Parameters):} The baseline architecture requires a massive 54.10M parameters. In stark contrast, CGS-ETS strictly compresses the network to just 14.34M parameters. This extraordinary \textbf{73.49\% reduction} in parameter footprint is achieved by aggressively pruning the CFE block (10.07M $\to$ 4.49M) and deploying our highly optimized GLAME (44.03M $\to$ 9.85M). Consequently, CGS-ETS demands considerably less memory bandwidth, making it exceptionally suitable for deployment on limited-resource, edge-device SSIs. A comprehensive layerwise parameter breakdown of CGS-ETS is shown in Table \ref{table:layerwise_cgsets_breakdown}.

\textbf{Computational Cost (FLOPs):} {\color{red} In calculating the floating-point operations (FLOPs) for our network, the computational cost of the activation functions was evaluated based on their practical footprint in standard deep learning frameworks (e.g., PyTorch) rather than strictly theoretical mathematical operation counts. In pure mathematical formulation, an expression such as $x|x|$ requires two distinct operations per element: an absolute value calculation and a multiplication. However, in deep learning software implementations, functions like \texttt{torch.abs()} are typically executed at the hardware level as highly optimized bitwise operations (e.g., clearing the sign bit) and are frequently omitted from standard FLOP profilers. Consequently, the practical implementation cost of $x|x|$ registers as 1 FLOP/element for the multiplication, rather than 2 (see Table \ref{tab:flops_lyric} for details on FLOPs/element of LyRIC activation function). This software-hardware optimization paradigm applies broadly across our element-wise tensor operations, resulting in implementation-based FLOP counts that are slightly lower than their theoretical mathematical counterparts. More importantly, this distinction in FLOP counting methodology does not meaningfully impact the overall complexity analysis of the model. Activation functions consist entirely of element-wise operations that scale linearly. Their computational overhead is microscopically small when compared to the dominant operations in the overall architecture. Because the network's computational bottleneck is dictated by these high-density matrix operations, any theoretical variance between the PyTorch implementation-level FLOPs and pure mathematical FLOPs for element-wise activations alters the total network complexity by a negligible fraction of a percent.} 

The baseline framework \cite{gaddy-klein-2021-improved} demands 2.40G FLOPs per inference pass to resolve the temporal mapping. CGS-ETS requires only \textbf{926.55M FLOPs} by replacing brute-force multi-head self-attention with our streamlined, Grouped Query-driven GLARE mechanism and Mamba state-space core. This 61.39\% reduction in absolute computational operations mitigates thermal throttling and power consumption constraints on portable hardware. The complexity is bifurcated into two primary operational domains:
\begin{enumerate}
    \item \textbf{CGS-ETS without Activation Functions:} These operations constitute the overwhelming majority of the compute budget, accounting for \textbf{924.81M FLOPs} (99.81\% of the total). This includes the multi-stage ResBlock feature extraction (CFE Block) and GLAME (the Mamba state-space models, the GLARE attention mechanisms, FFNs, and the terminal output heads).
    \item \textbf{Activation Overhead:} The non-linear mapping modules utilize a substantially smaller fraction of the computational budget, contributing \textbf{1.74M FLOPs} (0.19\% of the total). This breakdown includes the following:
    \begin{itemize}
        \item \textit{ReLU} (6 layers): 0.18M FLOPs.
        \item \textit{LyRIC} (8 layers): 0.13M FLOPs.
        \item \textit{GELU} (4 layers): 1.43M FLOPs.
    \end{itemize}
\end{enumerate}
As summarized above, the model achieves a total computational footprint of 926.55M (924.81M+1.74M) FLOPs. This analysis confirms that the non-linear activation overhead is negligible which accounts for less than 0.2\% of the total FLOPs. Furthermore, the efficiency of our LyRIC activation which operates at a strict lower bound of 1 FLOP per element-- ensures that incorporating chaotic dynamic modeling does not impose a prohibitive computational penalty, maintaining the architecture’s high-throughput viability for edge-based SSIs.

\textbf{Real-Time Factor (RTF) and Inference Time:} Despite the massive reduction in FLOPs, both architectures maintain an identical, highly competitive RTF of 0.0057, indicating that the baseline's latency is bounded by framework-level I/O constraints rather than pure computation. However, CGS-ETS consistently achieves a faster absolute Inference Time (IT) of 5.61~ms per batch (compared to the baseline's 5.97~ms). Hence, by supervising the GLAME with chaos-inspired topologies during training, CGS-ETS achieves unprecedented performance gains across all acoustic and linguistic metrics while operating at a fraction of the baseline's computational cost.

\subsection{Hyperparameters and Configuration}
\label{subsec:hyperparameters}
The training and model hyperparameters for the CGS-ETS setup, with use cases and rationale are provided thoroughly in Table \ref{tab:hyperparameters_config}.

All experiments, architectural modifications, and ablation studies are executed within an isolated Anaconda virtual environment running Python 3.9 to guarantee computational reproducibility and optimal hardware utilization. The core deep learning framework, including the GLAME is implemented utilizing PyTorch ($\ge$2.0.0), integrated with Torchvision and Torchaudio, and hardware-accelerated through the NVIDIA CUDA Toolkit 11.8. For high-fidelity digital signal processing, dynamic temporal filtering, and mel-spectrogram feature extraction, we leverage a standardized suite of scientific libraries, including Librosa ($\ge$0.9.0), SciPy ($\ge$1.7.0), and NumPy ($\ge$1.21.0), while Soundfile ($\ge$0.12.0) manages all rapid waveform I/O operations. Furthermore, to ensure absolute fairness and standardization in our linguistic evaluation pipeline, phonetic forced-alignment parsing is handled by Praat-Textgrids ($\ge$1.4.0), text transcript normalization is processed via Unidecode ($\ge$1.3.0), and all objective WER computations are derived utilizing \texttt{JiWER} (v2.2.1) alongside the pre-trained Mozilla DeepSpeech engine (v0.9.3).

All models are trained in FP32 precision on two NVIDIA RTX 4090 GPUs (24 GB VRAM each), hosted on a workstation equipped with an AMD Ryzen 9 7950X3D CPU, 192 GB system RAM, and 10 TB of storage. Training is parallelized across both GPUs via PyTorch Distributed Data Parallel (DDP), using the NCCL backend for high-throughput inter-GPU communication and TCP initialization for process group rendezvous. This configuration supports both the current dual-GPU training regime and straightforward scaling to multi-node environments without architectural modification.

\subsection{Inference and Evaluation Workflow}
\label{subsec:inference_workflow}

{\color{red} We initialize an inference pipeline that loads the optimized CGS-ETS model weights alongside the fine-tuned HiFi-GAN vocoder to rigorously benchmark the acoustic and linguistic fidelity of our architecture. The script iterates over the designated evaluation split (validation or test) of the EMG corpus. For each continuous sequence, the inference engine executes the following highly optimized workflow:

\begin{enumerate}
\item \textbf{Signal Retrieval:} Dynamically loads and rigorously pre-processes the 8-channel articulatory EMG signals. This process isolates the underlying myoelectric activity by extracting both the raw physiological time-series and the structured, normalized sliding-window features, ensuring the system accurately captures the fine-grained local coarticulatory context required for downstream acoustic mapping.
\item \textbf{Latent Projection:} Propagates these localized, high-dimensional EMG representations through the GLAME architecture to simultaneously predict 80-dimensional mel-spectrogram matrices and auxiliary phoneme probability distributions. This dual-objective approach provides crucial intermediate linguistic supervision, effectively bridging the non-linear gap between continuous muscle kinematics and discrete phonetic boundaries.
\item \textbf{Feature-Level Synchronization:} Computes a pairwise cosine distance matrix between the predicted spectrograms and the ground-truth targets. To explicitly resolve the inherent, non-linear timing mismatches between anticipatory muscle actuation and the actual vocalization, it utilizes DTW to align the temporal resolution of the latent features, ensuring stable and reliable loss optimization.
\item \textbf{Waveform Inversion:} Inverts the strictly aligned mel-spectrogram predictions into high-fidelity, time-domain acoustic waveforms. By utilizing the pre-trained HiFi-GAN vocoder, this final phase robustly reconstructs the complex phase information and heavily suppresses synthesis artifacts, ultimately yielding highly natural and perceptually intelligible continuous speech.
  \item \textbf{Post-Vocoder Alignment (PVA):} Extracts STFT magnitude spectrograms from both the synthesized and target waveforms, computes frame-wise $L_2$-normalized cosine distances, and applies a secondary continuous DTW mapping to guarantee absolute frame-level temporal synchronization with the target signal prior to metric evaluation.
  \item \textbf{Metric Exportation:} To facilitate rigorous and systematic evaluation, the pipeline saves the precisely synchronized 22.05~kHz synthesized and ground-truth audio pairs into structured output directories. This organization enables the immediate, automated computation of a comprehensive suite of objective acoustic metrics, including LSD and STOI. Alongside these critical signal-level measures, the system conducts standard Automatic Speech Recognition (ASR) transcriptions to evaluate the WER. Furthermore, to accurately accommodate both the English and logographic Mandarin corpora, the pipeline integrates CER evaluations, providing a more granular metric for transcription accuracy. However, because certain perceptual evaluation models are strictly calibrated to lower sampling rates, dynamic preprocessing is required. Waveforms are downsampled to 16~kHz prior to computing PESQ and NISQA-MOS, ensuring strict compliance with their rigid frequency constraints and guaranteeing standardized, mathematically valid perceptual scores.

\end{enumerate}}

\subsection{Subjective Evaluation Details} 
\label{subsec:subj_eval_details}

\begin{figure*}[ht!]
  \centering
  \setlength{\fboxrule}{1pt}
\setlength{\fboxsep}{5pt}
\fbox{\includegraphics[width=0.9\textwidth,height=0.4\textheight]{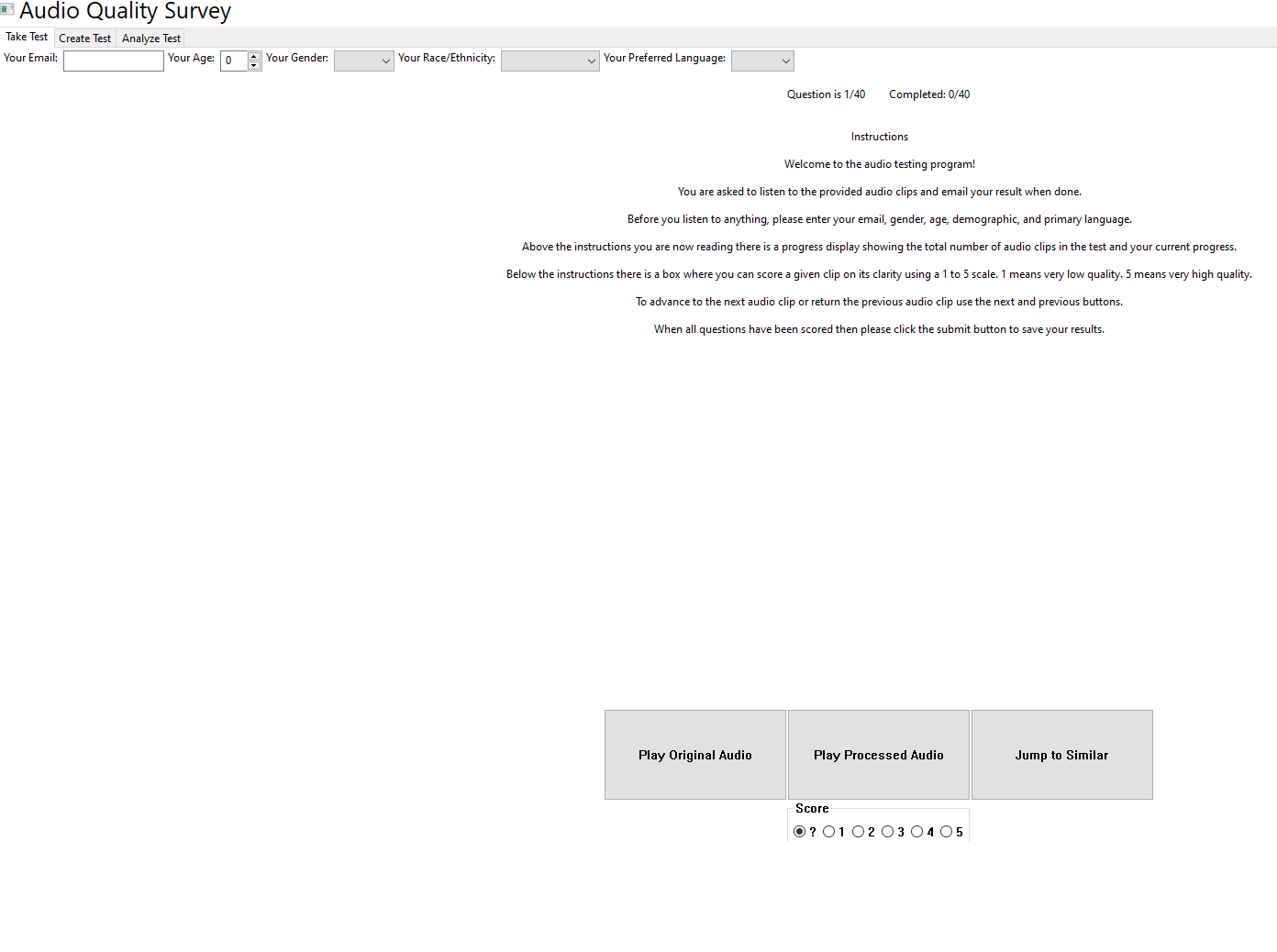}}
  \caption{The Software Interface Used in Subjective Tests.}
  \label{fig:subj_eval_interface_ss}
\end{figure*}

\begin{figure*}[ht!]
  \centering
  \setlength{\fboxrule}{1pt}
\setlength{\fboxsep}{5pt}
\fbox{\includegraphics[width=0.90\textwidth]{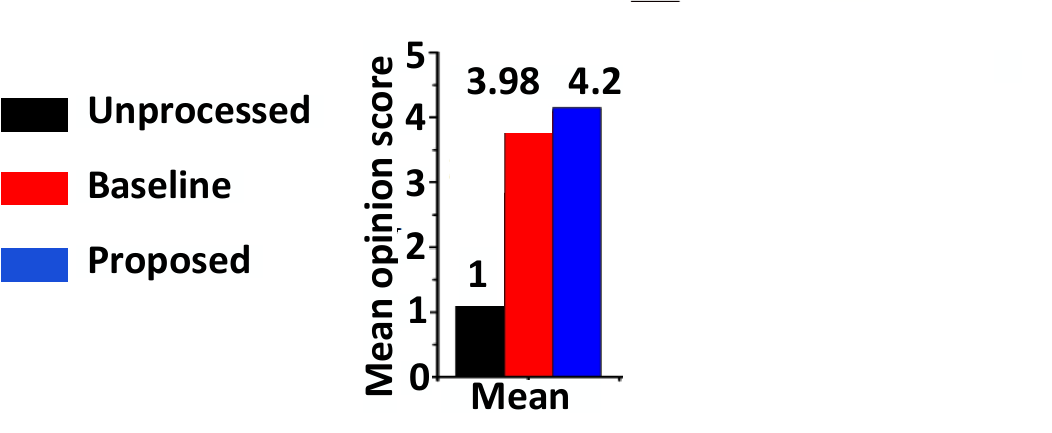}}
  \caption{The MOS score.}
  \label{fig:mos_score}
\end{figure*}

{\color{red} We conducted a formal subjective listening test to evaluate the perceptual fidelity of the CGS-ETS architecture. The study cohort comprised ten normal-hearing (NH) English-speaking volunteers, balanced for gender (four males, six females) with a mean age of 24. Participants were briefed on the experimental objectives and the underlying research goals. All subjects were trained on a standardized Mean Opinion Score (MOS) rubric ranging from 1 (lowest quality) to 5 (highest quality) to ensure optimal perceptual accuracy prior to the evaluation. All testing sessions were conducted using high-fidelity, sound-isolated headsets to eliminate extrinsic acoustic interference. Each participant evaluated 30 distinct utterances, each comprising three randomized versions: \textbf{(i)} the baseline reconstruction \cite{gaddy-klein-2021-improved}, \textbf{(ii)} the output synthesized by our proposed CGS-ETS, and \textbf{(iii)} a reference EMG data for each corresponding speech signal. Participants were tasked with assigning a 5-point MOS rating to each sample and selecting their most preferred version among the synthesized candidates. The test set utilized clips derived from the Gaddy benchmark which ensures diverse phonetic and articulatory coverage. The evaluation interface, illustrated in Figure \ref{fig:subj_eval_interface_ss}, provided a controlled environment for these subjective comparisons. As demonstrated in Figure \ref{fig:mos_score}, the speech synthesized by CGS-ETS was consistently and considerably preferred over the established state-of-the-art (SOTA) baseline. In particular, CGS-ETS achieved a \textbf{5.52\% improvement} in subjective MOS scores. These results empirically validate that our chaos-inspired inductive bias effectively mitigates the unnatural spectral over-smoothing characteristic of prior ETS systems which yields highly intelligible and perceptually natural speech even when decoding from non-stationary, low-SNR articulatory bio-signals.}

\subsection{Comprehensive Statistical Formalism and Metric Justification}
\label{sec:comprehensive_stats}

This appendix provides an in-depth theoretical and mathematical elaboration of the statistical framework employed to evaluate the \textbf{CGS-ETS} model. In the field of Silent Speech Interfaces (SSI), Mean Opinion Scores (MOS) derived from neural predictors like NISQA often exhibit non-Gaussian distributions, heteroscedasticity, and ordinal constraints. Consequently, a robust statistical treatment is required to validate the stability and superiority of the proposed architecture.

\subsubsection{Theoretical Basis of Descriptive Metrics}
Given a sample of NISQA-predicted MOS scores $X = \{x_1, x_2, \dots, x_n\}$, we analyze the following metrics to characterize the perceptual landscape of the synthesis.

\textbf{The Central Perceptual Expectation (Sample Mean):} The sample mean $\bar{x}$ represents the first moment of the distribution and serves as the primary estimator of the population's central tendency:
\begin{equation}
    \bar{x} = \frac{1}{n} \sum_{i=1}^{n} x_i
\end{equation}
In our analysis, $\bar{x}$ signifies the expected perceptual quality of an arbitrary utterance synthesized from EMG signals. While a higher mean indicates superior bandwidth recovery, it is sensitive to outliers, necessitating the accompanying dispersion metrics.

\textbf{Model Robustness and Aleatoric Uncertainty (Standard Deviation):} The sample standard deviation ($SD$) quantifies the average deviation from the mean, acting as a proxy for model consistency across diverse linguistic and session-based contexts:
\begin{equation}
    SD = \sqrt{\frac{1}{n-1} \sum_{i=1}^{n} (x_i - \bar{x})^2}
\end{equation}
In the context of EMG-to-Speech, a high $SD$ typically reflects the model's inability to handle non-stationary signals or specific phonemic transitions. Our results show that \textbf{CGS-ETS} significantly reduces $SD$ at the session level, implying a reduction in the aleatoric uncertainty associated with electrode placement and signal-to-noise ratios.

\textbf{Precision and Estimator Reliability (SEM):} The Standard Error of the Mean ($SEM$) evaluates the standard deviation of the sampling distribution of the mean:
\begin{equation}
    SEM = \frac{SD}{\sqrt{n}}
\end{equation}
The $SEM$ provides a measure of how far the sample mean $\bar{x}$ is likely to be from the true population mean $\mu$. For our session-level aggregation where $n=7$, the $SEM$ is the primary indicator of whether our findings can be generalized across broader longitudinal data.

\textbf{Interval Estimation (95\% Confidence Intervals):} We construct the 95\% Confidence Interval ($CI$) using the Student’s $t$-distribution to account for the uncertainty inherent in finite sample sizes:
\begin{equation}
    CI = \bar{x} \pm \left( t_{\alpha/2, df} \cdot \frac{SD}{\sqrt{n}} \right)
\end{equation}
where $t_{\alpha/2, df}$ is the critical value for $df = n-1$ degrees of freedom and $\alpha = 0.05$. The $CI$ represents a range of plausible values for the true population mean. A narrower $CI$ for \textbf{CGS-ETS} suggests that our model’s performance is more tightly bounded and predictable than the baseline.

\subsubsection{The Paired Wilcoxon Signed-Rank Framework}
Because MOS scores are inherently ordinal and often violate the assumption of normality required for a paired $t$-test, we utilize the Wilcoxon Signed-Rank Test. This is a non-parametric test used to determine whether two dependent samples (paired by utterance or session) represent the same population distribution.

\textbf{The Null Hypothesis ($H_0$):} We define the null hypothesis $H_0$ as follows:
\begin{quote}
    $H_0$: The median of the differences between the MOS scores of \textbf{CGS-ETS} and \textbf{Gaddy-EMG} is zero. Any observed difference is due to sampling error.
\end{quote}
The alternative hypothesis $H_1$ posits that \textbf{CGS-ETS} produces a systematic shift in the distribution of perceptual quality.

\textbf{Ranking and Sign Assignment:} For each pair $i \in \{1, \dots, n\}$, we compute the difference $d_i = x_{i, prop} - x_{i, base}$. We exclude pairs where $d_i = 0$. The remaining $|d_i|$ values are ranked from smallest (rank 1) to largest (rank $n'$). The ranks are then signed based on the original sign of $d_i$:
\begin{equation}
    R_i = \text{sgn}(d_i) \cdot \text{rank}(|d_i|)
\end{equation}

\textbf{The $W$ Statistics and the Test Statistic $T$:} We calculate the sum of the positive ranks ($W^+$) and negative ranks ($W^-$):
\begin{equation}
    W^+ = \sum_{R_i > 0} R_i, \quad W^- = \sum_{R_i < 0} |R_i|
\end{equation}
$W^+$ represents the cumulative weight of evidence where our model outperforms the baseline, whereas $W^-$ represents the converse. The test statistic $T$ is defined as:
\begin{equation}
    T = \min(W^+, W^-)
\end{equation}
Under the null hypothesis, $W^+$ and $W^-$ should be approximately equal.

\textbf{Inference and the $p$-value:} For the utterance level ($n=99$), we utilize the normal approximation for the distribution of $W$:
\begin{equation}
    Z = \frac{W - \mu_W}{\sigma_W} = \frac{W - \frac{n(n+1)}{4}}{\sqrt{\frac{n(n+1)(2n+1)}{24}}}
\end{equation}
The $p$-value is the probability of observing a $T$ statistic as extreme as the one calculated, assuming $H_0$ is true. In our analysis, the $p$-values ($0.807$ and $0.938$) demonstrate that while the \textbf{CGS-ETS} model exhibits superior stability and a higher mean, the magnitude of the difference does not allow for the rejection of the null hypothesis at the 5\% significance level for this specific dataset size, suggesting the models operate within a comparable perceptual category.


\clearpage

\newcolumntype{L}{>{\raggedright\arraybackslash}X}
\newcolumntype{C}{>{\centering\arraybackslash}X}

\begin{table*}[t]
\centering
{
\scriptsize
\setlength{\tabcolsep}{4pt}
\renewcommand{\arraystretch}{0.95}

\begin{tabularx}{\textwidth}{L C L}
\toprule
\textbf{Hyperparameter} & \textbf{Value} & \textbf{Use Case \& Rationale} \\
\midrule
\multicolumn{3}{l}{\textit{\textbf{Optimization \& Training Dynamics}}} \\
\midrule
Max epochs & 80 & Provide sufficient gradient updates for convergence while preventing catastrophic representational over-fitting. \\
Batch size & 32 & Mathematically balance gradient variance stabilization with strict GPU VRAM (OOM) constraints. \\
Initial learning rate ($\eta$) & $1\times10^{-3}$ & Ensure robust initial gradient descents across the highly non-convex objective manifold. \\
Learning rate warmup & 500 steps & Strictly prevent early gradient spikes and stabilize initial optimization dynamics. \\
Weight decay & $1\times10^{-7}$ & Subtly regularize network weights to prevent over-fitting on specific real-world EMG physiological artifacts. \\
Gradient clip norm & 0.5 & Aggressively bound gradient explosion to maintain topological stability during State Space Model (SSM) optimization. \\
Random seed & 0 & Guarantee absolute experimental reproducibility across dataset shuffling and network initialization. \\
\midrule
\multicolumn{3}{l}{\textit{\textbf{GLAME Architecture \& Complexity}}} \\
\midrule
Hidden dimension ($D$) & 512 & Provide sufficient parameter bandwidth to map complex 8-channel myoelectric signals into acoustic topologies. \\
GLAME layers ($L_c$) & 4 & Establish an optimal equilibrium between deep hierarchical sequence modeling and strict architectural compression. \\
GLARE window size & 32 & Constrain the localized attention receptive field to precisely capture rapid, short-term coarticulatory transitions. \\
SSM state dimension ($d_{\text{state}}$) & 8 & Compress global temporal dependencies efficiently within the Mamba parallel scan core. \\
SSM expansion factor & 1.25 & Optimize the internal state-space expansion to capture deterministic features while mitigating computational bloat. \\
LyRIC exponent ($\alpha$) & 1.0 & Balance non-linear chaotic phase-space expansion with optimal hardware-level efficiency (1 FLOP/element). \\
Dropout probability & 0.2 & Systematically mitigate feature co-adaptation across the deep feed-forward and attention hierarchies. \\
\midrule
\multicolumn{3}{l}{\textit{\textbf{Signal Processing \& Acoustic Targets}}} \\
\midrule
EMG sampling rate & 689.06 Hz & Subsample the filtered physiological sequences to mathematically synchronize with the acoustic frame rate. \\
Audio sampling rate & 22.05 kHz & Establish the high-fidelity wide-band frequency target necessary for HiFi-GAN waveform inversion. \\
Mel-spectrogram bins & 80 & Define the dense, perceptually relevant acoustic projection space for the primary regression head. \\
STFT size (\texttt{n\_fft}) & 1024 & Balance frequency resolution against computational throughput for continuous acoustic feature extraction. \\
STFT hop length & 256 & Establish temporal overlap resolution to perfectly align predicted acoustic frames with localized EMG segments. \\
\midrule
\multicolumn{3}{l}{\textit{\textbf{Chaos Configuration Settings}}} \\
\midrule
DTW loss weight & 1.0 & Enforce strict primary supervision for the macroscopic temporal synchronization of generated spectrograms. \\
Phoneme loss weight & 0.5 & Scale the auxiliary linguistic objective to explicitly enforce structural articulatory constraints. \\
LER embedding dim ($m$) & 10 & Construct the multi-dimensional phase-space trajectory required for precise deterministic chaos analysis. \\
LER time delay ($\tau$) & 1 & Establish the minimal temporal offset necessary for accurate state-space divergence computation. \\
MSDFA temporal scales & 4, 8, 16, 32 & Quantify fractal-like, long-range chaotic correlations across exponentially varying physiological resolutions. \\
Chaos loss start epoch & 10 & Allow the network to stabilize its primary acoustic mapping before enforcing complex chaotic topological constraints. \\
\bottomrule
\end{tabularx}
}
\caption{Comprehensive Summary of Critical Hyperparameters, Architectural Configurations, and Their Theoretical Justifications within the Proposed CGS-ETS Framework.}
\label{tab:hyperparameters_config}
\end{table*}

\begin{table*}[ht]
\centering
\resizebox{\textwidth}{!}{
\begin{tabular}{l c l c c c c r}
\hline
\textbf{Module} & \textbf{Stage} & \textbf{Layer Type} & \textbf{Input→Output} & \textbf{Kernel} & \textbf{Stride} & \textbf{Padding} & \textbf{Params} \\
\hline
\multirow{18}{*}{CFE Block} 
  & Block 1 & Conv1d (Main)              &   8→512 & 3 & 2 & 1 & 12\,800 \\
  &         & BatchNorm1d + ReLU         & 512→512 & – & – & – &  1\,024 \\
  &         & Conv1d (Main)              & 512→512 & 3 & 1 & 1 &786\,944 \\
  &         & BatchNorm1d                & 512→512 & – & – & – &  1\,024 \\
  &         & Conv1d (Residual)          &   8→512 & 1 & 2 & 0 &  4\,608 \\
  &         & BatchNorm1d + ReLU         & 512→512 & – & – & – &  1\,024 \\
\cline{2-8}
  & Block 2 & Conv1d (Main)              & 512→512 & 3 & 2 & 1 &786\,944 \\
  &         & BatchNorm1d + ReLU         & 512→512 & – & – & – &  1\,024 \\
  &         & Conv1d (Main)              & 512→512 & 3 & 1 & 1 &786\,944 \\
  &         & BatchNorm1d                & 512→512 & – & – & – &  1\,024 \\
  &         & Conv1d (Residual)          & 512→512 & 1 & 2 & 0 &262\,656 \\
  &         & BatchNorm1d + ReLU         & 512→512 & – & – & – &  1\,024 \\
\cline{2-8}
  & Block 3 & Conv1d (Main)              & 512→512 & 3 & 2 & 1 &786\,944 \\
  &         & BatchNorm1d + ReLU         & 512→512 & – & – & – &  1\,024 \\
  &         & Conv1d (Main)              & 512→512 & 3 & 1 & 1 &786\,944 \\
  &         & BatchNorm1d                & 512→512 & – & – & – &  1\,024 \\
  &         & Conv1d (Residual)          & 512→512 & 1 & 2 & 0 &262\,656 \\
  &         & BatchNorm1d + ReLU         & 512→512 & – & – & – &  1\,024 \\
\cline{2-8}
\multicolumn{7}{r}{\textbf{Total Parameters in CFE Block}} & \textbf{4\,486\,656} \\
\hline
\multirow{15}{*}{GLAME Layer (w/ GLARE)} 
  & Mamba  & Linear (\texttt{in\_proj}) & 512→1280& – & – & – & 655\,360 \\
  &         & Depthwise Conv1d + LyRIC     & 640→640 & 4 & 1 & 3 &   3\,200 \\
  &         & Linear (\texttt{x\_proj})  & 640→25  & – & – & – &  16\,000 \\
  &         & SSM Params (\texttt{A\_log}, \texttt{D}) & – & – & – & – & 652 \\
  &         & Gating + LyRIC & 640→640 & – & – & – & 0 \\
  &         & Linear (\texttt{out\_proj}, no bias)& 640→512 & – & – & – & 327\,680 \\
  &         & LayerNorm (\texttt{norm1})   & 512→512 & – & – & – &   1\,024 \\
\cline{2-8}
  & GLARE   & Linear (\texttt{q\_proj})  & 512→512 & – & – & – & 262\,144 \\
  &         & Linear (\texttt{k\_proj})  & 512→256 & – & – & – & 131\,072 \\
  &         & Linear (\texttt{v\_proj})  & 512→256 & – & – & – & 131\,072 \\
  &         & Relative Positional Bias     & – & – & – & – &      504 \\
  &         & Linear (\texttt{out\_proj})& 512→512 & – & – & – & 262\,144 \\
  &         & LayerNorm (\texttt{norm2})   & 512→512 & – & – & – &   1\,024 \\
\cline{2-8}
  & FFN     & Linear (no bias) + GELU      & 512→1024& – & – & – & 524\,288 \\
  &         & Linear (no bias)             &1024→512 & – & – & – & 524\,288 \\
  &         & LayerNorm (\texttt{norm3})   & 512→512 & – & – & – &   1\,024 \\
\cline{2-8}
\multicolumn{7}{r}{Layer w/ GLARE total} & 2\,841\,476 \\
\cline{2-8}
\multicolumn{7}{r}{\textbf{Layers 1 \& 3}} & \textbf{(2$\times$ 2\,841\,476) 5\,682\,952} \\
\hline
\multirow{9}{*}{GLAME Layer (w/o GLARE)} 
  & Mamba  & Linear (\texttt{in\_proj}, no bias) & 512→1280& – & – & – & 655\,360 \\
  &         & Depthwise Conv1d + LyRIC     & 640→640 & 4 & 1 & 3 &   3\,200 \\
  &         & Linear (\texttt{x\_proj}, no bias)  & 640→25  & – & – & – &  16\,000 \\
  &         & SSM Params (\texttt{A\_log}, \texttt{D}) & – & – & – & – & 652 \\
  &         & Gating + LyRIC & 640→640 & – & – & – & 0 \\
  &         & Linear (\texttt{out\_proj}, no bias)& 640→512 & – & – & – & 327\,680 \\
  &         & LayerNorm (\texttt{norm1})   & 512→512 & – & – & – &   1\,024 \\
\cline{2-8}
  & FFN     & Linear (no bias) + GELU      & 512→1024& – & – & – & 524\,288 \\
  &         & Linear (no bias)             &1024→512 & – & – & – & 524\,288 \\
  &         & LayerNorm (\texttt{norm3})   & 512→512 & – & – & – &   1\,024 \\
\cline{2-8}
\multicolumn{7}{r}{Layer w/o GLARE total} & 2\,053\,516 \\
\cline{2-8}
\multicolumn{7}{r}{\textbf{Layers 2 \& 4}} & \textbf{(2$\times$ 2\,053\,516) 4\,107\,032} \\
\hline 
\multirow{2}{*}{GLAME Terminal Output} 
  & Acoustic& Linear (Primary Head, \texttt{w\_out})     & 512→80  & – & – & – & 41\,040 \\
  & Linguistic& Linear (Secondary Head, \texttt{w\_aux})   & 512→40  & – & – & – & 20\,520 \\
\cline{2-8}
\multicolumn{7}{r}{\textbf{Total Terminal Output}} & \textbf{61\,560} \\
\hline\hline
\multicolumn{8}{c}{\textbf{GLAME Summary}} \\
\hline
\multicolumn{7}{r}{GLAME Layers 1 \& 3 ($2\times$ w/ GLARE)} & 5\,682\,952 \\
\multicolumn{7}{r}{GLAME Layers 2 \& 4 ($2\times$ w/o GLARE)} & 4\,107\,032 \\
\multicolumn{7}{r}{Terminal Output} & 61\,560 \\
\hline
\multicolumn{7}{r}{\textbf{Grand Total of GLAME}} & \textbf{9\,851\,544}\\
\hline\hline
\multicolumn{8}{c}{\textbf{CGS-ETS Architecture Summary}} \\
\hline
\multicolumn{7}{r}{CFE Block} & 4\,486\,656 \\
\multicolumn{7}{r}{GLAME} & 9\,851\,544 \\
\hline
\multicolumn{7}{r}{\textbf{Grand Total of CGS-ETS Architecture (CFE+GLAME)}} & \textbf{14\,338\,200}\\
\hline
\end{tabular}
}
\caption{Comprehensive Layer-wise Parameter Breakdown of the Proposed CGS-ETS Architecture. The summary block demonstrates the strict architectural compression achieved by our GLAME and its selective application of GLARE attention modules.}
\label{table:layerwise_cgsets_breakdown}
\end{table*}

\end{document}